\documentclass[prx,twocolumn,aps,epsf,showpacs,superscriptaddress,longbibliography,nofootinbib,notitlepage]{revtex4-2}

\usepackage[dvipsnames]{xcolor} 
\usepackage{tikz} 
\usepackage{graphicx}
\usepackage{dcolumn}
\usepackage{bm}
\usepackage{bbm}
\usepackage{dsfont}
\usepackage{physics}
\usepackage{multirow}
\usepackage{hyperref}
\usepackage{placeins} 
\usepackage[normalem]{ulem}
\usepackage{MnSymbol,comment}

\newcommand{\e}{\varepsilon}

\renewcommand{\>}{\rangle}

\newcommand{\Z}{\mathbbm{Z}}

\begin{document}

\preprint{APS/123-QED}

\title{Entanglement renormalization circuits for $2d$ Gaussian Fermion States}
\author{Sing Lam Wong}
\affiliation{Department of Physics and Astronomy, and Stewart Blusson Quantum Matter Institute, University of British Columbia, Vancouver, BC, Canada V6T 1Z1}
\author{Andrew C. Potter}
\affiliation{Department of Physics and Astronomy, and Stewart Blusson Quantum Matter Institute, University of British Columbia, Vancouver, BC, Canada V6T 1Z1}
\affiliation{Quantinuum, 303 S. Technology Ct, Broomfield, Colorado 80021, USA}

\begin{abstract}
The simulation of entangled ground-states of quantum materials remains challenging for classical computational methods in more than one spatial dimension, and is a prime target for quantum computational advantage. 
To this end, an important goal is to identify efficient quantum state preparation protocols that minimize the physical qubit number and circuit depth resources required to capture higher-dimensional quantum correlations. This work introduces a quantum circuit compression algorithm for Gaussian fermion states based on the multi-scale entanglement renormalization ansatz (MERA), which provides an exponential reduction in the circuit depth required to approximate highly-entangled ground-states relevant for quantum materials simulations. The algorithm, termed two-dimensional Gaussian MERA ($2d$ GMERA), extends MERA techniques to compress higher-dimensional Gaussian states. Through numerical simulations of the Haldane model on a honeycomb lattice, the method is shown to accurately capture area-law entangled states including topologically trivial insulators, Chern insulators, and critical Dirac semimetals. While Gaussian states alone are classically simulable, this approach establishes empirical upper bounds on quantum resources needed to prepare free fermion states that are adiabatically connected to correlated ground states, providing guidance for implementing these protocols on near-term quantum devices and offering a foundation for simulating more complex quantum materials. Finally, we develop a novel fermion-to-qubit encoding scheme, based on an expanding $2d$ topological order, that enables implementing fermionic rotations via qubit Pauli rotations with constant Pauli weight independent of system size. 
\end{abstract}
\maketitle 

\section{Introduction}
The simulation of entangled ground states of quantum materials in more than one spatial dimension remains challenging for classical computational methods, and is a prime target for quantum computational advantage.
However, practical limitations of quantum computing hardware motivate the need for efficient quantum state preparation protocols that minimize the physical qubit number and circuit depth resources required to capture higher-dimensional quantum correlations.

In this work, we construct an algorithm to approximately compress area-law-entangled two-dimensional (2d) Gaussian (noninteracting) fermion states into a quantum circuit taking a form similar to multiscale entanglement renormalization ansatz (MERA) tensor networks. 
Through numerical simulations, we empirically find that this Gaussian MERA (GMERA) approach affords an exponential reduction in circuit depth to prepare area-law entangled ground-states compared to general-purpose techniques to prepare an arbitrary (potentially volume-law) Gaussian fermion state.
In 2d, area-law Gaussian fermion states include both long-range entangled topological states (Chern-insulators), and critical states (nodal semimetals) that cannot be captured efficiently by $2d$ projected-entangled pair states (PEPS).

Since Gaussian fermion states can be efficiently described classically through their correlation matrices (a structure we exploit in constructing the GMERA algorithm), they, alone are not sufficient for demonstrating quantum advantage. However, there are multiple contexts in which Gaussian fermions state preparation can be a useful subroutine in quantum algorithms for simulating more general correlated states:
Many variational algorithms such as the unitary coupled-cluster ansatz~\cite{aanand2022a}, start with a mean-field (Hartree-Fock) ground-state, and variationally build in correlations. This approach can dramatically reduce the number of variational parameters (and associated classical optimization complexity) required to accurately approximate interacting ground-states of correlated electrons~\cite{niu2022holographic}.
Moreover, for gapped phases without intrinsic topological order, Gaussian fermion states are adiabatically connected by a finite-depth circuit to interacting ground-states~\cite{Hastings_2005}. This class includes long-range entangled states such as Chern insulators that are necessarily separated from short-range entangled product states by a phase transition, that adds a polynomial in system size overhead to adiabatic state preparation. The GMERA algorithm augmented with adiabatic evolution, can be used to establish (empirical) upper bounds on the qubit and gate resources required to simulated correlated topological insulators.

Our approach extends the $1d$ Gaussian matrix product state GMPS and GMERA techniques of Fishman and White (FW)~\cite{fishman2015compression} to 2d. 
The core of the FW algorithm involves decomposing the correlation matrix into spatially local blocks, identifying a set of orbitals within each block that are approximately unentangled from neighboring blocks, and \emph{distilling} these orbitals out by applying a (Gaussian) unitary transformation to disentangling them from the rest of the system and localize them onto a single site. 
Orbitals in each block that are not distilled, act as \emph{couriers} of entanglement between the blocks.
The distillation procedure is repeated on a sequence of partially overlapping sequence of blocks until all of the courier modes are also distilled.
Importantly, each Gaussian disentangling unitary acting on a block of $B$ sites can be compiled into $O(B^2)$ two-body fermionic rotations. When implementing these rotations on a qubit-based quantum processor using fermion-to-qubit encoding (F2QE), each fermionic rotation requires qubit strings with length scaling at most as the block radius $r$, adding an additional factor of $r$ overhead. Therefore, the complete quantum circuit implementation requires $O(B^2r)$ two-qubit gates.

As we explain, a direct generalization of these $1d$ methods to higher-dimension fails due to an uncontrolled buildup of increasingly long-range entanglement in complex higher-dimensional correlations among the courier modes as the entanglement renormalization steps are iterated.
To circumvent this issue we introduce a Wannierization procedure to prevent the development of high-dimensional correlations among the courier modes.
We remark, that our approach can be viewed as a strictly local, circuitized version of Zipper entanglement renormalization (ZER)~\cite{wong2022zipper}, a procedure for disentangling Gaussian states via quasi-local unitary operation generated by a sequence of continuous time Hamiltonian simulations, but with important modifications to ensure that it can be implemented as a quantum circuit. 

The $2d$ GMERA approach can be adapted to any lattice (or even non-translation invariant systems). For concreteness we demonstrate the approach numerically for the Haldane model ~\cite{haldane1998model} on a $2d$ Honeycomb lattice, which is relevant for a variety of $2d$ materials including graphene and transition metal dichalcogenides (TMDs), and whose phase diagram encompasses a rich set of phases including trivial- , topological- (Chern insulator), and quantum cirtical (Dirac semimetal) states. These three classes of states display distinct patterns of entanglement: Despite having short-range correlations and an energy gap, the Chern insulator presents a topological obstruction to preparation by a short-depth circuit~\cite{dubail2015tensor}, and also cannot be represented by a self-similar MERA~\cite{Li2019entanglement}. The Dirac semimetal state is even more highly-entangled, exhibiting quasi-long-range (algebraically decaying) critical correlations and entanglement.
We find empirically, that all of these area-law entangled ground states in an $L\times L$ size system can be efficiently approximated by a GMERA built from spatial blocks $B$ of radius $r$ with approximation error $\epsilon$ decaying exponentially in $r$. Translating these numerical scaling forms into quantum circuit resources, this implies that approximating these states in an $L \times L$ size system with error $\epsilon$ can be achieved by a fermionic circuit depth $D\approx \log L *B^2 = \log L *\log^4(1/\epsilon)$ without considering the fermion-to-qubit encoding (F2QE). When implementing this circuit on a qubit-based processor, standard $2d$ F2QE's introduce an additional factor of $L$ overhead in the circuit depth. However, by expanding a $\Z_2$ topological order, we can reduce the overhead from $L$ to $r$, leading to a total qubit circuit depth scales as $D \approx \log L * \log^5(1/\epsilon)$.
In terms of number of qubits, adapting the method of ~\cite{anad2023holograph}, the $2d$ GMERA can be recasted into a \emph{sequential circuit}, which, using qubit reset and reuse techniques~\cite{schon2005sequential,kim2017holographic,kim2017noiseresilient,barratt2021parallel,foss-feig2021holographic} enables a reduction in the number of qubits required from $q\sim O(L^2)$ to: $q\sim O(L\log L * \log^2(1/\epsilon))$~\footnote{Recasting as a sequential circuit incurs a multiplicative $\times L$ increase in circuit depth, but without changing the total gate count. This trade off is beneficial if qubit resources are limited.}


\section{Compressing a correlation matrix into a quantum circuit}
Gaussian fermion states are completely captured by their two-point correlations, characterized by the correlation matrix
\begin{align} 
C_{ij} = \langle \hat{c}^{\dagger}_i \hat{c}_j \rangle.
\end{align}
where $c_i^\dagger$ creates a fermion at local orbital $i$ (which we henceforth refer to as a ``site" though, in general, the local orbital labels could run over different spin- and orbital- flavors on a spatial site).

An $n\times n$ correlation matrix for a Gaussian pure state can be diagonalized by a Gaussian unitary transformation:
\begin{align}
    C = w \Xi w^\dagger, ~~~
    \Xi = \mathbbm{1}_{k} \oplus 0_{n-k}
\end{align}
where $w$ is a unitary matrix that rotates between onsite and eigenorbital bases, $\Xi$ is a diagonal matrix of occupation numbers of each eigen-orbital, and  $k$ is the total particle number.~\footnote{Here, we assume without loss of generality, that the correlation matrix has a $U(1)$ number conservation symmetry (superconducting systems can be captured by generalizing to the Nambu spinors $\Psi = \begin{pmatrix} c_i & c_i^\dagger\end{pmatrix}$ or using a Majorana representation~\cite{Majorana1937}).} 
The filled and empty blocks have $k$- and $n-k$ fold degeneracy, and there are many suitable choices of bases for these bands.
The strategy of correlation matrix compression is to find an approximately local set of basis states for the filled and empty bands respectively, and to find a local quantum circuit to approximate the many-body unitary $W \equiv \exp[c^\dagger_i (\log w)_{ij} c^{\vphantom\dagger}_j]$ that rotates the product state $|1\rangle ^{\otimes k} |0\rangle ^{\otimes n-k}$ into the desired Gaussian state with correlation matrix $C$, where $|1\rangle (|0\rangle)$ denote filled (empty) Fock states.

FW introduced a procedure~\cite{fishman2015compression} to compress a $1d$ Gaussian correlation matrix into a MPS or MERA format that can be respectively implemented by sequential or tree-like  quantum circuits.
We begin by reviewing this technique, and discuss why a direct extension of this approach to $2d$ fails, due to an uncontrolled build up of correlations and entanglement, which motivates our modified $2d$ GMERA ansatz.

\subsection{$1d$ GMPS and GMERA from correlation matrix}
\begin{figure*}[t]
    \centering
    \includegraphics[width=1 \textwidth]{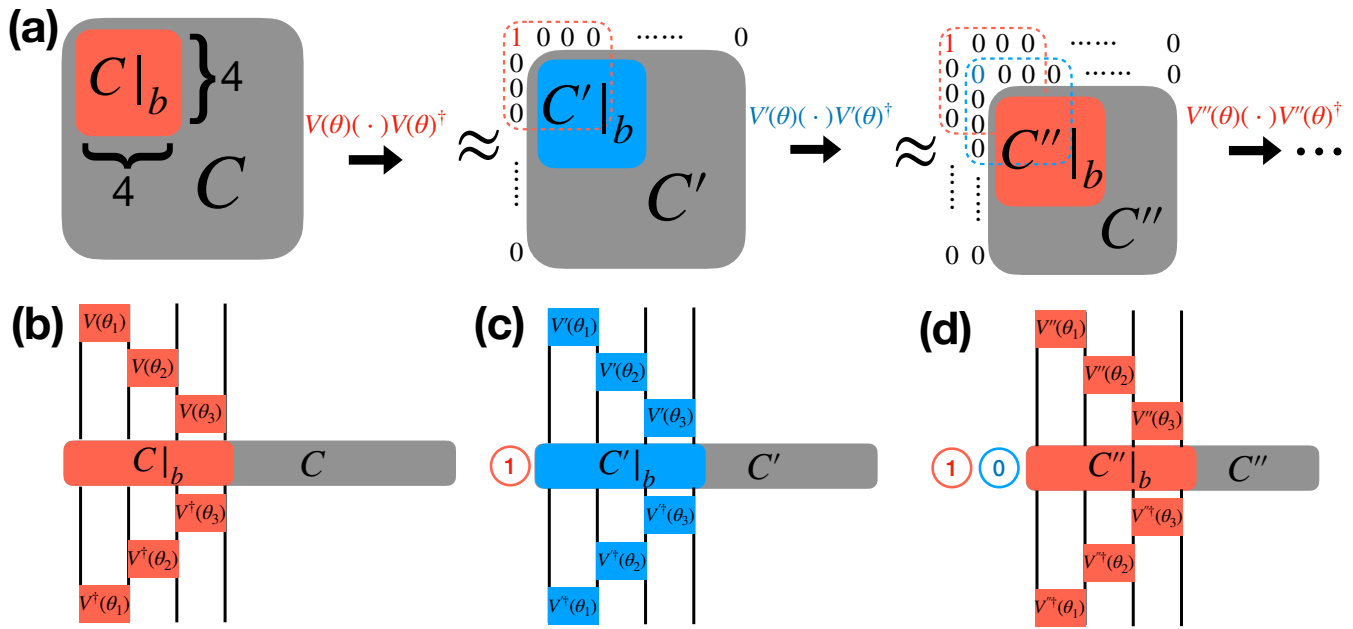}
    \caption{Illustration of the $1d$ GMPS compression procedure. (a) Schematic representation of the iterative block diagonalization process, where successive applications of unitary transformations $V(\theta)$ isolate local modes with occupation numbers near 1 (filled) or 0 (empty). Starting with a correlation matrix $C$, blocks of size 4 are analyzed sequentially. (b) The first transformation $V(\theta) = V_3(\theta)V_2(\theta)V_1(\theta)$ composed of three nearest-neighbor rotation gates acts to isolate a filled mode by rotating the eigenvector of the block correlation matrix $C|_b$ with eigenvalue closest to 1 to the first site, resulting in the filled mode marked by the red circled '1' in (c). (c) After shifting to the next block, a second transformation $V'(\theta)$ isolates an empty mode, leading to the empty mode marked by the blue circled '0' in (d). (d) The procedure continues iteratively until all modes are classified and isolated. This decomposition provides an efficient compression of the correlation matrix into a sequence of $2\times2$ nearest-neighbor rotation gates that can be implemented as a quantum circuit. The accuracy of the compression is controlled by how close the eigenvalues of the block correlation matrices are to 0 or 1, which is determined by the block size and the entanglement structure of the state.}
    \label{fig:1d_GMPS_procedure}
\end{figure*}

The $1d$ GMPS \cite{fishman2015compression} and GMERA procedure begins by examining the correlation matrix restricted to a sub-block of $B$ nearby sites, $C|_{B}$. Diagonalizing $C|_{B}$ yields a set of sub-block eigenvector, eigenvalue pairs: $\phi_\alpha,0\leq \eta_\alpha \leq 1$.
The degree to which block eigenorbital $\alpha$ is localized within the block is characterized by the Shannon entropy of its eigenvalue:
\begin{align}
    S(\eta) = -\eta\log\eta - (1-\eta)\log(1-\eta).
\end{align}
For a correlation matrix representing a locally-entangled state such as a ground-state of a local Hamiltonian, for a sufficiently large radius $r$, some eigenvalues will be very close to 0 or 1, corresponding to $S(\eta) \approx 0$, due to the limited entanglement structure.
These eigenmodes, collectively referred to as \emph{frozen modes}, as their occupation number has almost no quantum fluctuations, and they are essentially uncorrelated with the rest of the system outside the block in question.

Through a sequence of nearest-neighbor rotation gates, one can transform the basis to isolate these modes at the left boundary of the block. 
These rotations provide a local-circuit approximation to the the unitary that rotates an eigen-orbital of the full correlation matrix $C$ onto a single site, with the accuracy of the approximation controlled by the inter-block entanglement $S(\eta)$.
The remaining modes with intermediate eigenvalues that are not close to $0,1$ are termed \emph{courier modes} as they carry the quantum correlations between different blocks cannot be accurately disentangled by local operations acting within the block.

\paragraph{$1d$ GMPS:} In the $1d$ GMPS procedure, after disentangling the frozen modes to left boundary of the block, one then translates the block to the next $B$ entangled modes, and repeats the above procedure.
Iterating this procedure sequentially disentangles the system from left to right.
The resulting sequential circuit approximately disentangles the correlated state into a product of nearly-unentangled filled- and empty orbitals (See Fig.~\ref{fig:1d_GMPS_procedure}).
Conversely, applying the inverse of this circuit to a product of filed and empty orbitals approximately prepares the correlated ground-state.

\begin{figure}[h]
\centering
\includegraphics[width=0.5 \textwidth]{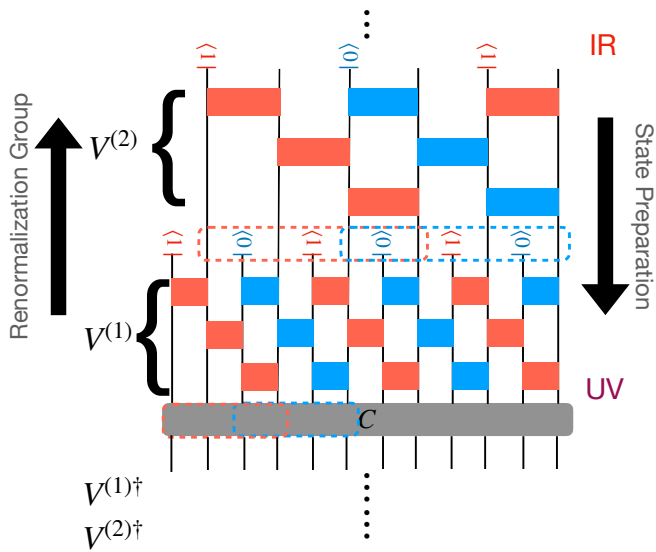}
\caption{Dual interpretation of the $1d$ GMERA procedure. The circuit diagram shows how a correlation matrix C can be processed in two complementary directions. Reading from bottom to top (UV to IR) represents a renormalization group (RG) flow, where V(1) and V(2) are successive layers of local 2×2 unitary rotations that systematically disentangle degrees of freedom from shorter to longer length scales. The red/blue coloring indicates filled/empty modes being disentangled, with dashed boxes highlighting the increasing length scales of each RG step. Alternatively, reading from top to bottom (IR to UV) represents a state preparation circuit, where the inverse transformations progressively entangle product states of filled ($|1\rangle$) and empty ($|0\rangle$) modes to build up the desired correlated state.}
\label{fig:1d_GMERA_procedure}
\end{figure}

\paragraph{$1d$ GMERA:}
FW also introduce a related recursive procedure to construct a Gaussian MERA (GMERA) circuit that performs the disentangling in an RG-like fashion.
In the $1d$ GMERA procedure, instead of sequentially disentangling orbitals a series of overlapping blocks, one instead disentangles orbitals from a layer of non-overlapping blocks in parallel.
This results in a coarse-grained chain with fewer degrees of freedom. Recursively applying this coarse-graining procedure gives rise to an RG-like circuit that approximately disentangles the correlated Gaussian fermion state.
This disentangling circuit can be viewed as a MERA tensor network consisting of alternating unitaries (the rotations that disentangle an orbital from a block), and isometries (that project out the disentangled orbitals.
Alternatively, the inverse of the disentangling circuit applied to a product state of filled and empty inputs can approximately prepare the desired correlated state (See Fig.~\ref{fig:1d_GMERA_procedure}).

\subsection{Memory efficient implementation with qubit reuse}
A particularly attractive feature of states prepared by sequential or MERA-type circuits is that their observables can be sampled using using a small number of qubits by leveraging qubit recycling techniques to opportunistically measuring qubits that have completed their role in the circuit, and resetting them to reuse in other parts of the circuit.
For the $1d$ GMPS procedure, qubit reuse enables implementation with number of qubits scaling as the block size $B$ independent of system size~\cite{fossfeig2024}. Moreover, $1d$ MERA circuits (including GMERA) can be recast as seqential circuits, and implemented with number of physical qubits scaling  as  $\sim \log(L)$, i.e. logarithmically in the length of the system, $L$.
These techniques can dramatically increase the size and complexity of systems that can be modeled with a given amount of quantum memory.

\subsection{Challenges of correlation matrix compression in 2d}
While the GMPS and GMERA methods work well in quasi-$1d$ geometries, $1d$ ground-states can generally be efficiently simulated by classical MPS methods, and quantum advantages on ground-state simulation problems likely reside in at least two spatial dimensions (as opposed to quantum dynamics simulation, which are classically difficult in any dimension). 
In the remainder of this work, we focus on generalizing the $1d$ GMERA methods to $2d$ GMERA methods.
To start, we illustrate why a direct generalization of the $1d$ techniques generally leads to an uncontrolled build up of entanglement causing the compression scheme to fail.
To address this issue, we next introduce an extra Wannierization procedure, to attempt to localize the entanglement produced within each local block during the GMERA procedure.

\begin{figure*}[t]
    \centering
    \hspace*{-0.2in}
    \includegraphics[width=1 \textwidth]{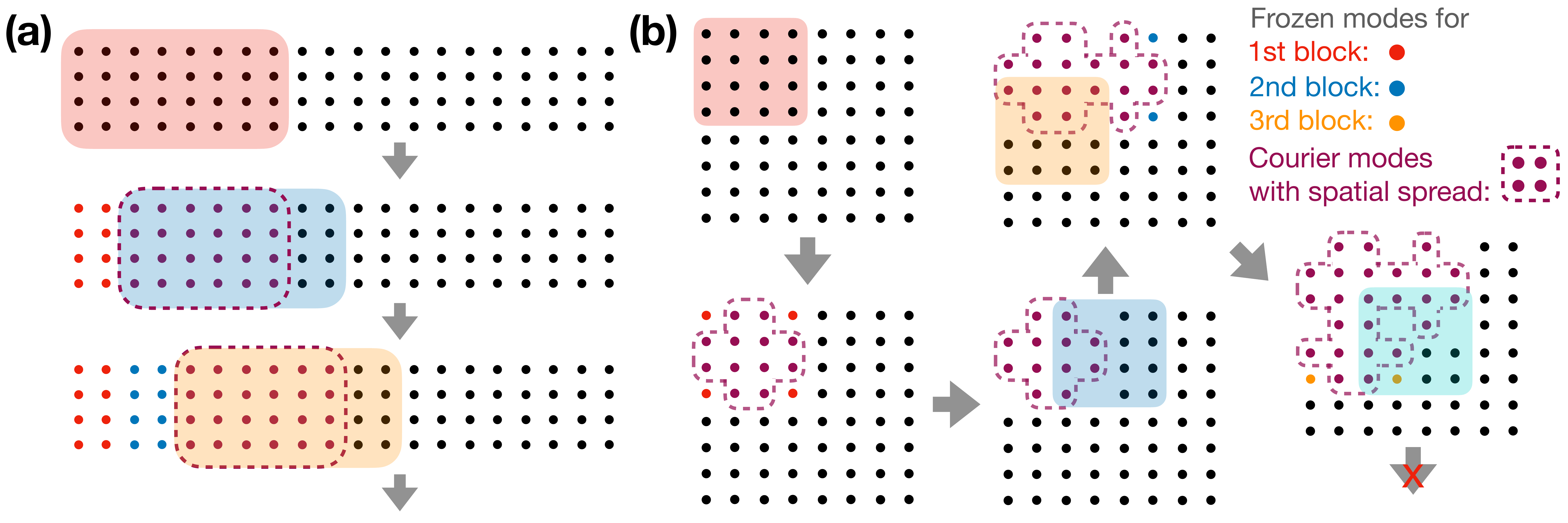}
    \caption{Comparison of GMPS implementation in quasi-$1d$ versus $2d$ systems. (a) Successful application in quasi-1d: The narrow width confines the spatial spread of courier modes between successive blocks, allowing complete coverage and effective sequential processing. Starting with the leftmost block (red), frozen modes (red) are rotated to the left boundary while courier modes ((purple dots with dashed outlines showing spatial spread)) remain confined within the system width. Subsequent blocks (blue, orange) can fully encompass and process all courier modes from previous steps due to the quasi-$1d$ geometry. (b) Failure in $2d$ systems: The procedure breaks down due to uncontrolled spread of courier modes. Starting with the first block, frozen modes are rotated to the block boundaries (red), but courier modes extend beyond the block boundaries. Subsequent blocks (blue, orange, cyan) attempt to rotate their frozen modes to their respective block boundaries, but show decreasing efficiency in this distillation as courier modes from previous steps spread beyond their boundaries. This fundamental problem persists regardless of whether we choose to rotate frozen modes to the block boundaries and courier modes to the center, or vice versa. 
    }
    \label{fig:Quasi-1d-&-2d-GMPS}
\end{figure*}

The applicability of Gaussian Matrix Product State (GMPS) methods can be best understood by first examining quasi-$1d$ systems, where the width of the system is much smaller than its length. In this geometry, when we process each block sequentially, the spatial spread of courier modes remains confined by the narrow width of the system. Crucially, each subsequent block completely encompasses the courier modes generated by the previous block as shown in Fig. \ref{fig:Quasi-1d-&-2d-GMPS}(a), allowing for consistent and effective mode distillation throughout the procedure. This complete coverage ensures that we can maintain control over the entanglement structure as we progress along the quasi-$1d$ system.

A straightforward generalization of this approach fails for $2d$ systems. 
As depicted in Fig.~\ref{fig:Quasi-1d-&-2d-GMPS}, distilling orbitals in one block create courier modes with extended entanglement, which cannot be removed by subsequent blocks. Iterating the block-distillation procedure then inevitably leads to uncontrolled build-up of entanglement and a failure of the GMPS procedure.
Below, we introduce a modified version of this scheme that introduces a Wannierization step to maintain the locality of courier modes to avoid this entanglement build-up. 
The modification is inspired by a Zipper Entanglement Renormalization procedure (ZER) \cite{wong2022zipper}, a recently-introduced holographic method that uses continuous-time Hamiltonian evolution to produce a quasi-local MERA-like unitary to disentangle Gaussian fermion states. 
Our $2d$ GMERA can be loosly thought of as a strictly local version of ZER that enables its implementation with gate-based quantum circuits.

\section{$2d$ GMERA Algorithm}
\begin{figure*}[t]
\centering
\includegraphics[width=\textwidth]{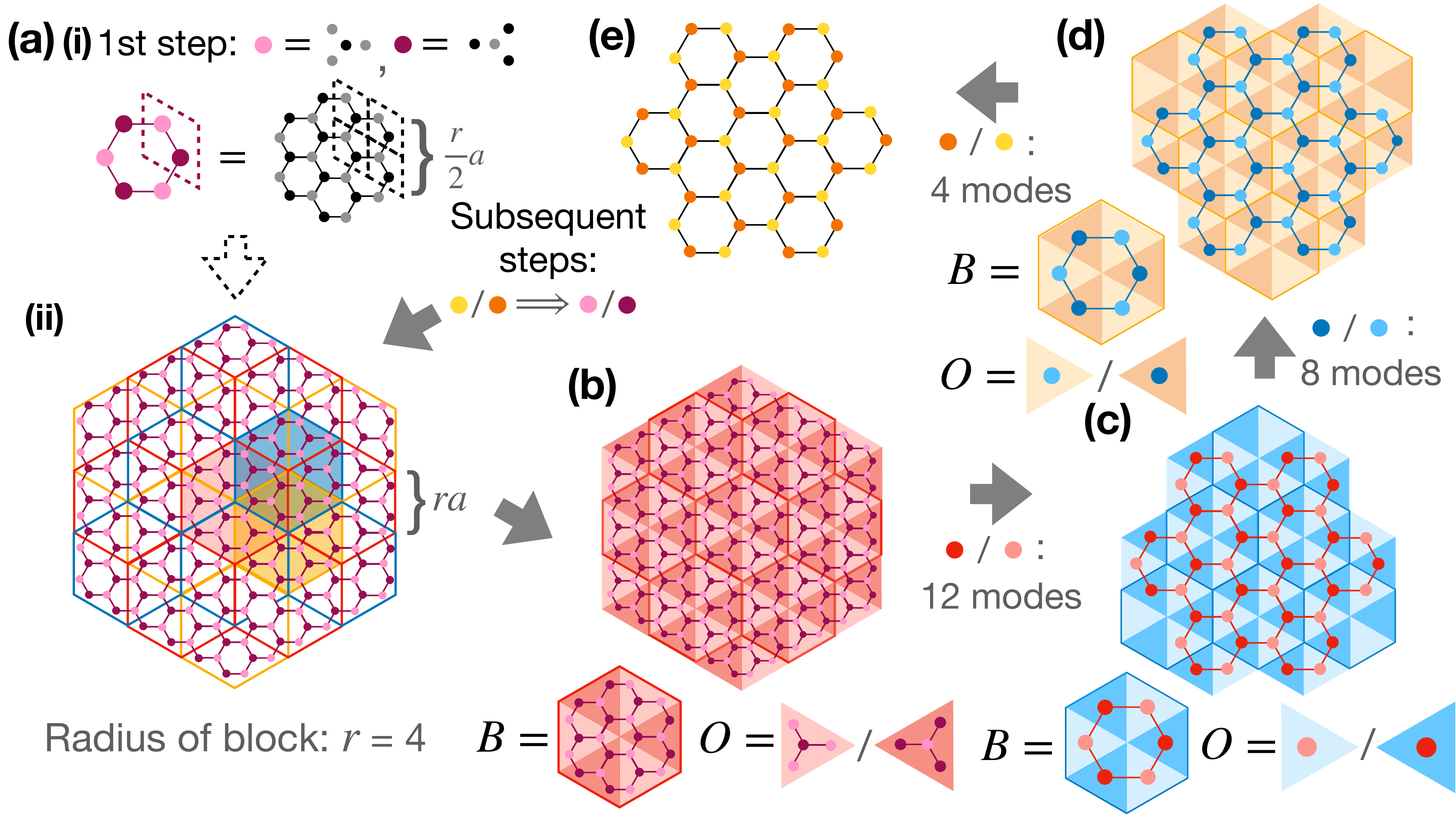}
\caption{2d GMERA procedure on the honeycomb lattice for $r=4$. (a)(i) First step initialization: Starting with an enlarged unit cell (represented in purple rhombus) of size $(r/2)a = 2a$ where $a$ is the lattice vector containing $2*2^2=8$ physical modes (represented in black rhombus). For honeycomb lattice, rhombus is the natural choice for unit cell, but due to the geometry of our distillation process, we pick the hexagon as our unit cell.(a)(ii) Implementation overview showing the three hexagonal blocks (red, blue, and green) representing the three sequential disentangling layers comprising one complete GMERA step. Each hexagonal block has radius $r = 4$ lattice vectors. (b) First disentangling layer: The hexagonal regions $B$ (red) are where we analyze the restricted correlation matrices. The triangular regions $O$ contain dots representing sites where we restrict our projector for finding the seeds of wannierization. These triangular regions also indicate where the coarse-grained degrees of freedom will be centered after wannierization, forming the honeycomb lattice shown in (c). After this disentangling step, each red dot (indicated near the arrow) represents 12 modes. (c) Second disentangling layer: The blue hexagonal blocks $B$ now process the coarse-grained lattice from the first step, with triangular regions $O$ serving as new wannierization centers. Each blue dot represents 8 modes in the resulting coarse-grained lattice. (d) Third disentangling layer: Green hexagonal blocks $B$ process the further coarse-grained lattice, with each green dot representing 4 modes. (e) After processing all three layers, the remaining courier modes form a coarse-grained honeycomb lattice with the same geometric structure as the original lattice but with expanded lattice constant. To continue the GMERA procedure, the orange/yellow dots play the role of purple dots in a new iteration of step (a)(ii), but now with a lattice constant twice the original size. }
\label{fig:GMERA-schematic}
\end{figure*}

The 2d GMERA algorithm provides a quantum circuit implementation for systematically disentangling free-fermion states through a sequence of strictly local operations.

\subsection{Implementation on the Honeycomb Lattice}
For concreteness, we illustrate the technique first for a honeycomb, as illustrated in Fig.~\ref{fig:GMERA-schematic}  for block radius $r=4$, and then comment on generalizations to a general $2d$ lattice.
The honeycomb lattice 2d GMERA procedure begins with an enlarged unit cell of size $\frac{r}{2}a\times \frac{r}{2}a$ where $a$ is the lattice vector and $r$ is the block radius (Fig.~\ref{fig:GMERA-schematic}(a)(i)). Each GMERA step consists of three sequential disentangling layers (Fig.~\ref{fig:GMERA-schematic}(a)(ii)), with each layer processing a set of non-overlapping hexagonal blocks that are shifted relative to each other.

The procedure for each disentangling layer works as follows:

\paragraph{Block Analysis} For each hexagonal block $B$ in the current layer (shown as hexagons in Fig.~\ref{fig:GMERA-schematic}(b-d)), we diagonalize the restricted correlation matrix $C|_B$ to identify modes that are approximately disentangled from the complement of the block $B$. These modes are strictly contained within the block:
    \begin{equation}
        C|_B = v^{(B)} \Lambda^{(B)} v^{(B)\dagger}
    \end{equation}
    where $v^{(B)}$ is a unitary matrix whose columns represent the eigenvectors, and $0\leq \Lambda^{(B)}\leq 1$ is a diagonal matrix of block eigenmode occupation numbers.
    Based on the eigenvalues in $\Lambda^{(B)}$, we classify the eigenvectors into three categories:
    \begin{itemize}
        \item Filled modes ($v^{(B)}_f$) with eigenvalues close to 1
        \item Empty modes ($v^{(B)}_e$) with eigenvalues close to 0
        \item Courier modes ($v^{(B)}_c$) with intermediate eigenvalues
    \end{itemize}
    The filled and empty modes represent degrees of freedom that can be distilled out, while the courier modes carry entanglement to the next scale.
    
    In our approach, each hexagonal block $B$ in Fig.~\ref{fig:GMERA-schematic}(b) contains 
    $8 \times r \times 3 = 24 r$ modes. 
    The least-entangled $6r$ of these modes are treated as frozen (approximately filled or empty) and distilled out, leaving $18r$ Courier modes for each block $B$.
    This $1/4$-distillation is repeated in each of the three disentangling layers, ultimately distilling out $3/4$ of all the modes in a complete GMERA step, leaving $6r$ remaining courier modes that must be dealt with in future GMERA steps. 

\paragraph{Wannierization of Courier Modes:} To prevent the uncontrolled buildup of entanglement as the GMERA RG proceeds, we build an approximately local basis for the courier modes near to the triangular regions $O$ formed at the triple intersections of hexagonal blocks from the three different disentangling layers (shown as shaded triangles in Fig.~\ref{fig:GMERA-schematic}(b-d)).

    The Wannierization process \cite{Marzari2012,Sakuma2013} involves several steps:
    \begin{enumerate}
        \item For each hexagonal block $B$, form the projector onto its courier modes:
        \begin{equation}
            P^{(B)}_c = v^{(B)}_c v^{(B)\dagger}_c
        \end{equation}
        
        \item Restrict this projector to each triangular overlapping region, $O$, and diagonalize it:
        \begin{equation}
            P^{(B,O)}_c = s^{(B,O)} \Gamma^{(B,O)} s^{(B,O)\dagger}
        \end{equation}
        where $\Gamma^{(B,O)} = \text{diag}(\lambda^{(B,O)}_1, \ldots, \lambda^{(B,O)}_n)$ contains eigenvalues arranged in descending order.
        
        \item For the first disentangling step in Fig.~\ref{fig:GMERA-schematic}(b), we have 
        $18 r$
        courier modes to distribute among 6 triangular regions $O$. Thus, for each triangular region, we select the first 
        $6r$
        eigenmodes $s^{(O)}_c$ correspond to $\Gamma^{(B,O)}$ as seeds for Wannierization.
        
        \item We formulate the Wannierization as an orthogonal Procrustes problem \cite{Gower2004}, seeking the optimal unitary transformation $w^{(B)}_c$ that best aligns the courier modes with the seeds identified in (c):
        \begin{equation}
            w^{(B)}_c = \arg\min_w \|v^{(B)}_c w - s^{(B)}_c\|_F
        \end{equation}
        where $s^{(B)}_c=\bigoplus_{O's}s^{(O)}_c$ combines all the seed vectors from the triangular regions associated with block $B$, and $\|\cdot\|_F$ denotes the Frobenius norm.
        
        This optimization has a closed-form solution: given the singular value decomposition:
        \begin{equation}
            \sigma^{(B)}_c 
            = t^{(B)}_c \Sigma^{(B)}_c (u^{(B)}_c)^\dagger,
        \end{equation}
        For this procedure to be well-defined, all singular values in $\Sigma^{(B)}_c$ must be strictly positive. This condition is expected to be satisfied in our case, as the frozen modes we distill out in each step are topologically trivial, ensuring no topological obstruction to the Wannierization process.
        The optimal transformation is:
        \begin{equation}
            w^{(B)}_c = u^{(B)}_c (t^{(B)}_c)^\dagger
        \end{equation}
        
        \item The resulting Wannierized courier modes $\tilde{v}^{(B)}_c = v^{(B)}_c w^{(B)}_c$ are orthonormal and maximally localized within the triangular overlapping regions, and at the same time ensuring strict locality in the hexagonal region B. These courier modes will be centered at the triangular regions, forming the sites of the coarse-grained lattice for the next step.
    \end{enumerate}

    \paragraph{Sequential Processing of Layers}: As shown in Fig.~\ref{fig:GMERA-schematic}(b-d) for the case of $r=4$, we sequentially process the three disentangling layers:
    \begin{itemize}
        \item The first layer (red, Fig.~\ref{fig:GMERA-schematic}(b)) processes hexagonal blocks containing physical modes. After Wannierization, each red dot centered at a triangular region represents $\frac{3}{4}\times\frac{24r}{6} = 12$ modes in the resulting coarse-grained lattice.
        
        \item The second layer (blue, Fig.~\ref{fig:GMERA-schematic}(c)) processes the coarse-grained lattice from the first step. The blue hexagonal blocks now contain the coarse-grained degrees of freedom from the first layer, and after Wannierization, each blue dot represents $\frac{2}{4}\times\frac{24r}{6} = 8$ modes in the newly formed coarse-grained lattice.
        
        \item The third layer (green, Fig.~\ref{fig:GMERA-schematic}(d)) processes the further coarse-grained lattice from the second step, with each orange dot representing $\frac{1}{4}\times\frac{24r}{6} = 4$ modes after Wannierization.
    \end{itemize}

After processing all three disentangling layers, we have completed one full GMERA step. The remaining courier modes form a coarse-grained honeycomb lattice with the same geometric structure (Fig.~\ref{fig:GMERA-schematic}(e)). The final coarse-grained system contains 4 modes per site, which is the same as the initialization step in (a)(i), making the procedure self-similar. We iterate the procedure by treating the orange dots as new physical modes (analogous to the purple dots in the first step) and then returning to step (a)(ii), but considering larger hexagonal regions in each subsequent iteration. This iterative approach allows us to systematically disentangle correlations at increasingly larger length scales.
A convenient feature of this construction is that it maintains a self-similar geometry for each RG step (though the disentangling operations at each RG are not necessarily scale invariant, but rather are dictated by the correlation structure of the Gaussian state being represented).

\subsection{Generalization to Arbitrary Lattices}

The procedure described above can be generalized to arbitrary lattices by adapting the following components:

\begin{enumerate}
    \item \textbf{Block Structure}: Define appropriate blocks with radius $r$ that cover the entire lattice. The number and arrangement of these blocks will depend on the lattice geometry.

    \item \textbf{Layer Organization}: Determine the minimum number of disentangling layers needed to ensure entanglement within the same length scale are considered. For the honeycomb lattice, three layers are sufficient, but other lattices may require different numbers.

    \item \textbf{Overlapping Regions}: Identify the junction points where blocks from different disentangling layers meet. These regions will serve as the centers for Wannierization of courier modes.

    \item \textbf{Coarse-Graining Structure}: Ensure that the coarse-grained lattice formed by the courier modes preserves the geometric structure of the original lattice, allowing for iterative application of the GMERA procedure.
\end{enumerate}

The key insight of the 2d GMERA approach is the use of Wannierization to maintain the locality of courier modes in the overlapping region among different disentanling steps, preventing the uncontrolled spread of entanglement that would otherwise occur in direct applications of 1d techniques to 2d systems. By processing blocks in layers and carefully localizing the courier modes to overlapping regions, we can systematically disentangle the state while preserving its local entanglement structure.

\begin{figure*}[t]
\centering
\includegraphics[width=\textwidth]{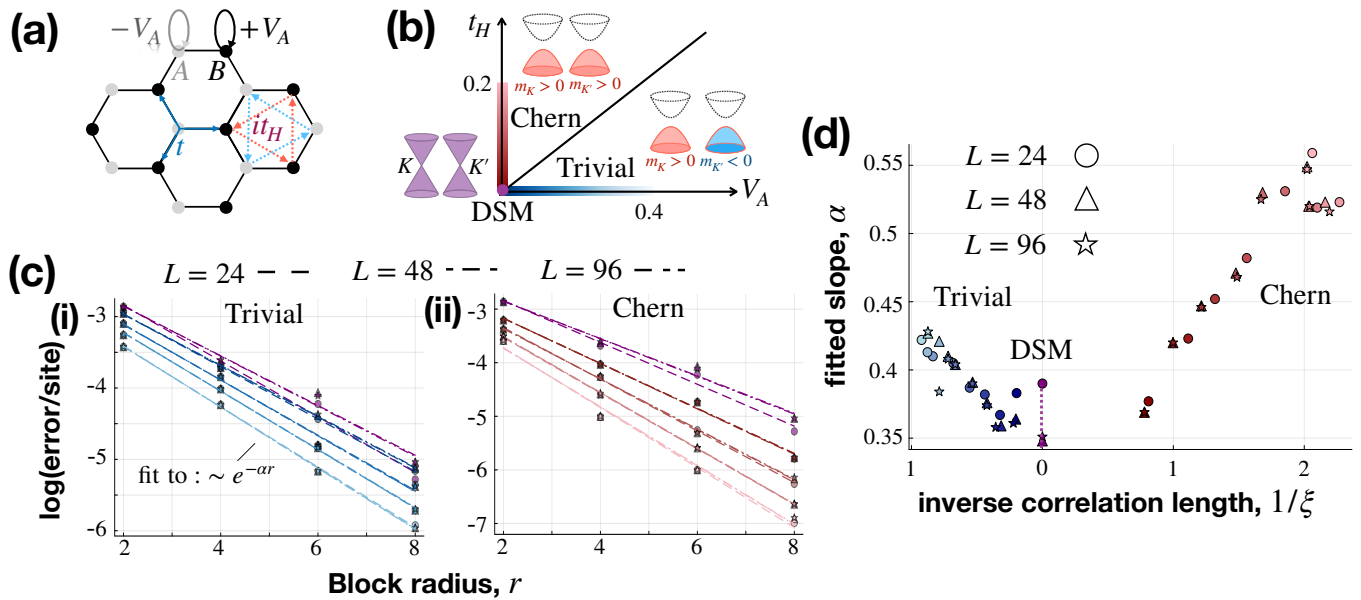}
\caption{{\bf $2d$ GMERA results for the Haldane model.}
(a) Schematic of the terms in the honeycomb Haldane model Eq.~\ref{eq:haldane}. A,B sublattice sites are shown in gray, black respectively. (b) Phase diagram of Eq.~\ref{eq:haldane}, including a schematic depiction of the (massless or massive Dirac) dispersion of the low-energy electronic states near the K and K' points in the Brillouin zone. Red and blue shading of the Dirac particles respectively indicate positive or negative $\frac{1}{2}$ contribution to the Chern number. The blue and red shaded line segments indicate the regions of parameter space explored in our numerical simulations, where at most one of $t_H$ or $V_A$ is non-zero. (c) Block-size scaling of the GMERA approximation error (logarithm of error per site) versus block radius $r$ for different system sizes $L = 24, 48$, and $96$. Data points from different system sizes collapse onto straight lines indicating exponential improvement with block size independent of system size. (i) For the trivial insulator case, the purple data points represent the Dirac semimetal ($V_A = 0$), while the blue points show trivial insulators with increasing $V_A$ values from $V_A = 0.1$ (darkest blue) to $V_A = 0.4$ (palest blue) in intervals of $0.1$. (ii) For the Chern insulator case, the purple data points again represent the Dirac semimetal ($t_H = 0$), while the red points show Chern insulators with increasing $t_H$ values from $t_H = 0.08$ (darkest red) to $t_H = 0.2$ (palest red) in intervals of $0.04$. The slopes of these lines become steeper (indicating faster convergence) as the correlation length decreases within both trivial and Chern phases. (d) The fitted exponential decay rate $\alpha$ from panel (c) plotted against the inverse correlation length $1/\xi$ for both trivial (blue points) and Chern insulator (red points) phases. For the trivial insulator (blue), $V_A$ ranges from $0.05$ to $0.4$ with $0.05$ intervals, while for the Chern insulator (red), $t_H$ ranges from $0.06$ to $0.2$ with $0.02$ intervals. The approximately linear relationship between $\alpha$ and $1/\xi$ demonstrates that states with shorter correlation lengths can be compressed more efficiently. Data from different system sizes collapse onto the same curves, with the critical point (purple) marking the transition between trivial and topological phases.
\label{fig:HaldaneModel}}
\end{figure*} 

\section{Numerical simulations}


\subsection{Honeycomb Haldane Model}
We demonstrate the $2d$ GMERA procedure for a  honeycomb Haldane model~\cite{haldane1998model} for a single spin species of electrons, whose phase diagram encompasses multiple paradigmatic phases including both trivial and topological (Chern) gapped insulators trivial insulator, separated by quantum critical Dirac semimetal:
\begin{align}
    H & = -t\sum_{\<ij\>} c^\dagger_i c^{\vphantom\dagger}_j
    - it_H \sum_{\llangle ij \rrangle} s_{ij}c^\dagger_i c^{\vphantom\dagger}_j
    -\frac 12V_A\sum_{i}(-1)^i c_i^\dagger c_i^{\vphantom\dagger}
    +{\rm h.c.}
    \label{eq:haldane}
\end{align}
where $c_i^\dagger$ creates an electron on site $i$, and the remaining parameters are detailed below. 

The nearest neighbor hopping, $t$, alone, gives a gapless Dirac semimetal (DSM) with two Dirac cones.
The DSM is a quantum critical point with long-range entanglement, and infinite correlation length corresponding to scale-invariant correlations that decay algebraically with distance.
These critical correlations and entanglement are expected to be captured by a (G)MERA wavefunction that is asymptotically self-similar.

Turning on a staggered potential, $V_A$, with opposite sign on the $A$ ($B$) sublattice, gives each Dirac cone opposite sign masses. This results in a topologically trivial insulator, with only short-range entanglement, that should be accurately described by a $2d$ PEPS wavefunction. Alternatively, we anticipate that this short-range entangled state can be accurately approximated by terminating the GMERA recursion after a finite number of RG steps.

When $V_A=0$, the (imaginary) second neighbor hopping, $t_H$, whose sign $s_{ij}=\pm 1$ is assigned based on whether $ij$ goes along or against the direction of the arrows shown in Fig.~\ref{fig:HaldaneModel}(a), gives the Dirac cones the same sign mass resulting in a Chern insulator with Chern number $C={\rm sign}(t_H)$.
Despite being short-range correlated, the Chern insulator has long-range entanglement and cannot be prepared from the trivial insulator by a short depth circuit.
Moreover, there is an obstacle to preparing a Chern insulator with a scale-invariant MERA~\cite{Li2019entanglement}, which produces either strictly vanishing or asymptotically infinite correlation length, in contrast to the Chern insulator's finite but non-vanishing correlation length.
Our $2d$ GMERA procedure bypasses this no-go constraint by breaking the self-similarity assumption. Though it is geometrically self-similar, preserving the structure of the honeycomb, the unitary operations are scale dependent (different for each RG step), being generated from the physical correlation matrix at each scale.

For general parameter values, Fig.~\ref{fig:HaldaneModel}(b) shows a schematic phase diagram. To isolate the physics of each representative phase, in the following numerical simulations, we focus our attention on the red and blue shaded line segments where at most one of $t_H$ or $V_A$ is non-zero.

\subsection{Numerical results and approximation error scaling}
To benchmark the performance of our GMERA algorithm, we analyze the error in approximating the correlation matrix as a function of block radius $r$ for different phases and system sizes. We define the (root-mean-square, RMS) error-per-site as:
\begin{align}
    \epsilon = \sqrt{
    \frac{1}{N}
    \sum_{i,j} (C_{ij} - C_{ij}^{\text{approx}})^2},
\end{align}
where $C_{ij}$ and $C_{ij}^{\text{approx}}$ are the exact and GMERA-approximated correlation matrices respectively, and $N$ is the total number of sites. As shown in Fig.~\ref{fig:HaldaneModel}(ci,ii), 
the error decreases exponentially with the block radius $r$, as $\epsilon \sim e^{-\alpha r}$ where the decay rate $\alpha$ depends on the parameter values $t_H,v_A$, but is independent of system size (as expected for short-range correlated states).

The effectiveness of GMERA in compressing states with different correlation lengths is evident in the systematic downward shift of these lines. For the trivial insulator phase (Fig.~\ref{fig:HaldaneModel}(ci)), increasing the on-site potential $V_A$ from $0$ to $0.4$ results in shorter correlation lengths and correspondingly lower error at fixed block radius. 
Similar behavior is observed in the Chern insulator (Fig.~\ref{fig:HaldaneModel}(cii)), though we remark that the correlation length exhibits a non-mononotonic dependence on $t_H$ for $t_H>0$ 
This behavior aligns with intuitive expectation that states with shorter correlation lengths require less quantum information to be carried by the courier modes between blocks, enabling more efficient compression at fixed block size.

To further understand the relationship between compression efficiency and correlation length, we analyze how the fitted slope $\alpha$ from the error scaling $\text{error-per-site} \approx \exp(-\alpha r)$ varies with the inverse correlation length $1/\xi$. As shown in Fig.~\ref{fig:HaldaneModel}(d), we plot $\alpha$ against $1/\xi$ for both trivial and Chern insulator phases, with the correlation length $\xi$ extracted from the exponential decay of the real-space correlation functions. For the trivial insulator phase (blue points, $V_A$ varying from $0.05$ to $0.4$), the magnitude of the slope $|\alpha|$ increases approximately linearly with $1/\xi$, indicating that states with shorter correlation lengths can be compressed more efficiently. A similar linear trend is observed in the Chern insulator phase though with a different slope (red points, $t_H$ varying from $0.06$ to $0.2$).
Crucially, we note that the parameter $\alpha$ remains finite at the gapless/critical DSM point ($V_A,t_H\rightarrow 0$), showing that the GMERA can efficiently capture scale-invariant patterns of long-range entanglement.

Finally, in all cases the decay rate, $\alpha$, shows little dependence on system size beyond $L\sim 24$. Though it is impossible to rule out very slow (e.g. logarithmic) drifts with $L$, with any finite size numerics, our data strongly suggests that $\alpha$ remains constant throughout the phase diagram in the thermodynamic limit ($L\rightarrow \infty$).


\subsection{From GMERA to GPEPS: Truncating the RG procedure}\label{GMERA-to-GPEPS}

\begin{figure*}[t]
\centering
\includegraphics[width=\textwidth]{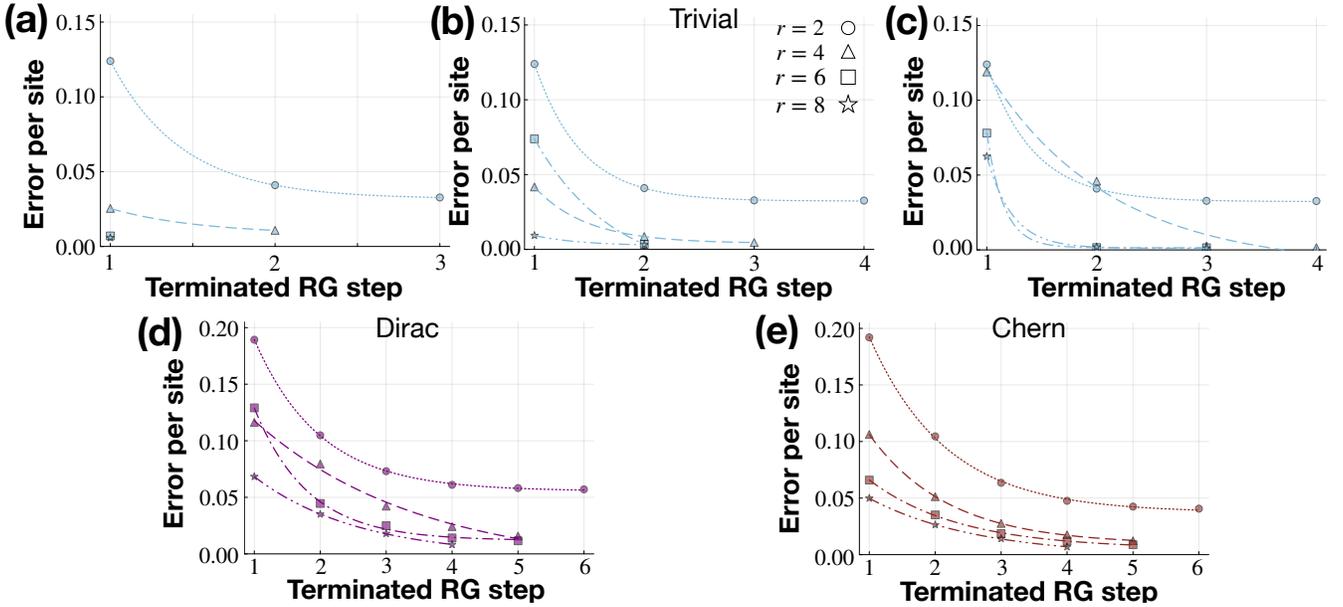}

\caption{\textbf{Thresholded GMERA results.} Trivial insulator phase with $V_A=0.4$ and correlation length $\xi=1.14$ using different threshold values: (a) $\zeta = 10^{-3}$, (b) $\zeta = 10^{-4}$, and (c) $\zeta = 10^{-5}$.In panel (a), with the largest threshold value, the RG procedure terminates extremely quickly—after just one step for block radii $r = 6$ and $r = 8$, with only $r = 2$ requiring three RG steps. As we decrease the threshold to $10^{-4}$ in panel (b), more RG steps are required before automatic termination, with improved accuracy compared to panel (a). For $r = 6$ and $r = 8$, the procedure terminates after two steps, while smaller block sizes require additional steps. In panel (c), with the smallest threshold of $10^{-5}$, we observe a similar trend with more stringent accuracy requirements, clearly showing error plateaus across different block sizes. The plateau behavior is especially pronounced for $r = 2$ (circles), where additional RG steps beyond the third provide minimal improvement in accuracy. Across all three threshold values, larger block sizes consistently yield better accuracy with fewer RG steps, confirming that trivial insulators can be efficiently represented by finite-depth circuits. (d) Dirac semimetal phase and (e) Chern insulator phase with $\zeta = 10^{-3}$. In striking contrast, both the Dirac semimetal and Chern insulator phases show no automatic termination even with the most lenient threshold of $10^{-3}$. These phases require progressively more RG steps to improve accuracy, with errors continuing to decrease through all available steps. This fundamental difference in behavior reveals the distinct entanglement structures of these phases: trivial insulators possess short-range entanglement that can be fully captured by finite-depth circuits, while the long-range entanglement in Dirac semimetals and Chern insulators requires the full GMERA procedure without termination. This confirms that topologically non-trivial and critical states cannot be efficiently represented by shallow quantum circuits.
\label{fig:GMERA-to-GPEPS-threshold}}
\end{figure*}

For short-range correlated states, it may not be necessary to continue the GMERA RG recursion indefinitely. Truncating the RG recursion of the GMERA procedure after a fixed number of RG steps (and distilling all remaining orbitals in the final step) generates a Gaussian PEPS (GPEPS). 
Examining the accuracy of this resulting GPEPS versus the number of RG steps before termination (Fig. \ref{fig:GMERA-to-GPEPS-threshold}) reveals fundamental distinctions between the entanglement structure in the trivial and topological phases, despite both phases having short-range correlations.

\subsubsection{Thresholded Distillation Approach}
For short-range entangled states,  it should not be required to iterate the RG procedure indefinitely to capture correlations. We introduce alternative scheme, which we refer to as ``thresholded distillation". In this approach, rather than distilling exactly 1/4 of the total modes in each local region, we greedily distill all modes whose eigenvalues fall within a specified threshold $\zeta$ of $0$ or $1$ (i.e., all modes with eigenvalues $\leq \zeta$ or $\geq 1-\zeta$)~\footnote{In practice, to preserve the self-similarity of the RG structure with hexagonal blocks, we demand that the number of distilled modes is divisible by $6$, and distill the largest number of sub-treshold modes compatible with this constraint.}
 For systems with particle-hole symmetry at half-filling, this specifically requires that both the number of filled and empty modes are equal and divisible by $3$. When the count of modes falling within our threshold is not exactly divisible by $6$, we include additional modes with eigenvalues closest to our threshold until this divisibility constraint is satisfied.

This thresholded approach can lead to automatic termination of the RG procedure. When all the remaining modes are classified as frozen at a given RG step, the RG flow terminates naturally as no courier modes remain to be processed.
This automatic termination serves as a distinctive signature that differentiates short-range entangled states from those with longer-range entanglement, for which termination does not occur until the RG has run its course and the entire system is coarse-grained into a single block.

\subsubsection{Distinct Behaviors Across Different Phases}
The results shown in Fig.~\ref{fig:GMERA-to-GPEPS-threshold} reveal striking differences between trivial insulators and other phases. To make a fair comparison between the trivial and chern insulator phases (comparable correlation length), we pick $V_A=0.4$ for trivial insulator with correlation length $\xi=1.14$ and small Haldane term $t_H=0.08$ for chern insulator with correlation length $\xi=1.00$.\break

\paragraph{Trivial Insulator:} 
For the trivial insulator phase (Fig.~\ref{fig:GMERA-to-GPEPS-threshold}(a-c)), we examine three different threshold values ($10^{-3}$, $10^{-4}$, and $10^{-5}$). Key observations include:

\begin{enumerate}
\item \textbf{Automatic termination}: The Thresholded approach terminates automatically when all modes fall within the threshold, indicating complete disentanglement within the specified accuracy.

\item \textbf{Plateau behavior}: Once the effective length scale exceeds the system’s correlation length, further renormalization steps do not significantly improve the accuracy of the state approximation. This is consistent with the short-range entangled nature of gapped, topologically trivial systems.

\item \textbf{Threshold dependence}: As expected, smaller thresholds ($10^{-5}$) enable more accurate approximations and can distinguish more subtle entanglement structures compared to larger thresholds ($10^{-3}$).
\end{enumerate}

In panel (a) of Fig.~\ref{fig:GMERA-to-GPEPS-threshold}, with the largest threshold value ($\zeta = 10^{-3}$), the RG procedure terminates extremely quickly—after just one step for block radii $r = 6$ and $r = 8$, with only $r = 2$ requiring three RG steps. As we decrease the threshold to $10^{-4}$ in panel (b), more RG steps are required before automatic termination, with improved accuracy. For $r = 6$ and $r = 8$, the procedure terminates after two steps, while smaller block sizes require additional steps.

In panel (c), with the smallest threshold of $10^{-5}$, we observe a similar trend with more stringent accuracy requirements, clearly showing error plateaus across different block sizes. The plateau behavior is especially pronounced for $r = 2$ (circles), where additional RG steps beyond the third provide minimal improvement in accuracy.

Across all three threshold values, larger block sizes consistently yield better accuracy with fewer RG steps, confirming that trivial insulators can be efficiently represented by finite-depth circuits.\break

\paragraph{Dirac Semimetal and Chern Insulator:}
In contrast to the trivial insulator phase, our Thresholded GMERA reveals fundamentally different behavior for both the Chern insulator and the Dirac semimetal, as shown in Fig.~\ref{fig:GMERA-to-GPEPS-threshold} (d-e). Unlike the trivial insulator phase, neither of these phases exhibits automatic termination of the RG flow, even with our most lenient threshold value of $\zeta = 10^{-3}$.

The DSM has scale-invariant (power-law decaying) correlations that cannot be accurately captured by a short-range correlated PEPS. This structure manifests in a continued, non-saturating reduction of error with increasing RG steps.

The Chern insulator represents an intermediate case. Despite having short-range correlations, the Chern insulator state(s) necessarily involve long-range entanglement, and there are fundamental obstructions to representing states with non-vanishing Chern number via any constant bond-dimension PEPS. As for the DSM, this long-range entanglement structure results in a non-saturating reduction of error with increasing RG steps.

These comparative results across different phases provide strong numerical evidence supporting our theoretical understanding of entanglement structures. The thresholded GMERA empirically demonstrates that: whereas short-range entangled states (trivial band-insulators) can be efficiently represented as finite-depth circuits and the thresholded RG procedure terminates after a constant number of iterations, long-range entangled critical or topological insulators present an obstacle to representation via shallow circuits signaled by a non-self-terminating GMERA.
This fundamental difference in behavior provides strong numerical evidence for the distinct entanglement structures of these phases. 

\section{From GMERA to circuit}
Read in reverse (from IR to UV), the GMERA procedure takes the form of a local quantum circuit that prepares an entangled fermionic state. In this section, we explain how this quantum circuit can be compiled into two-qubit gates that can be implemented on standard quantum processors. This proceeds in two stages, first we review how a unitary matrix acting on $B$ fermionic modes can be compiled into $O(B^2)$ two-body fermionic rotations of the form $e^{-i\theta_{ij} (c_i^\dagger c_j^{\vphantom\dagger}+h.c.)}$. Next, we discuss how these fermionic rotations can be encoded into a qubit-based processor using a novel fermion-to-qubit encoding (F2QE) scheme involving an expanding $2d$ topological encoding that is tailored to the $2d$ GMERA structure.

\subsection{Compiling block disentanglers}
A key step in implementing the $2d$ GMERA is efficiently compiling the block disentanglers acting on $B$ modes into sequences of two-body fermionic rotations. The block disentangler is a unitary matrix $\tilde{v}^{(i)} = (v^{(i)}_f | \tilde{v}^{(i)}_c | v^{(i)}_e)$ that separates the modes into filled, courier, and empty sectors. This $B\times B$ single-particle unitary can be decomposed into $O(B^2)$ $2\times 2$ single-particle rotations by rotating $\tilde{v}^{(i)}$ into identity column by column - each column requires at most $B$ $2\times 2$ rotations, leading to $O(B^2)$ rotations in total. Each $2\times 2$ rotation can then be converted into a two-body fermionic rotation of the form $e^{-i\theta_{ij}(c^\dagger_i c_j + \text{h.c.})}$ acting on a many-body fermionic Hilbert space. 


\subsection{Fermion-to-qubit encoding (F2QE)}
In a quantum processor built from fermionic modes, these fermionic rotations could be directly implemented. 
Implementing the Gaussian fermion rotations of the form $e^{-i\theta_{ij} (c_i^\dagger c_j^{\vphantom\dagger}+h.c.)}$ on a conventional, qubit-based quantum processor requires encoding the fermionic bilinears into qubits using a fermion-to-qubit encoding (F2QE) scheme. F2QE adds additional circuit complexity, which depends both on the encoding scheme, dimensionality, and connectivity of the fermion interactions. Below, we briefly reviewing the standard $1d$ F2QE based on the Jordan Wigner (JW) transformation \cite{Jordan1928berDP} and show that its overhead scales only with the block radius, $r$, and not the system size for for $1d$ GMPS and GMERA methods.

\paragraph{$1d$: Jordan Wigner F2QE: }
For $1d$ systems with local interactions, F2QE can be done by the standard Jordan-Wigner (JW) mapping, which converts fermionic bilinears $c_i^\dagger c_j^{\vphantom\dagger}$ into strings of qubit operators stretching from site $i$ to $j$.
In $1d$ GMPS circuits, the maximal Pauli weight of these JW strings is the linear block size, $r$, which makes the qubit gate count a multiplicative factor of $r$ larger than the number of free fermion rotations (which scales like $r^2$), leading to an overall gate count scaling as $G_{\rm GMPS} \sim r^3 L$.

For architectures with flexible (any-to-any) gate connections, such as trapped-ion and neutral atom platforms, the cost of fermion-to-qubit encoding via JW transformation for the $1d$ GMERA is similar. Naively, based on Fig.~\ref{fig:1d_GMERA_procedure}, unitaries acting on coarse-grained blocks (e.g. $V^{(2)}$ appear to involve longer-range connections in terms of the UV lattice. However, these long-range gates pass over the qubits injected to convert the isometries, which, at a given stage of the RG, have a definite fermion parity. Hence, the corresponding factor of the Pauli-$Z$ in the JW string for these sites can be replaced by its fermion parity eigenvalue, $\pm 1$, and these injected qubits do not increase the operator weight of fermion-to-qubit encoding compared to the GMPS~\cite{corboz2010simulation}. Notice, however, that in a fixed-connectivity architecture with two-qubit gates available only between neighboring qubits in a $1d$ geometrically local layout, there would be an additional sequence of SWAP gates required to implement the longer-range rotations at intermediate RG steps. These SWAP gates increase the gate count to implement the $1d$ GMERA circuit by a factor proportional to the UV length of the largest IR block size, which is proportional to the total system size $L$.

\paragraph{$2d$: Expanding $\Z_2$ topological order encoding: }
The JW encoding is poorly suited for $2d$ geometries: since it requires imposing a $1d$ ordering on a $2d$ lattice, certain short-range fermion bilinears become encoded into Pauli strings with Pauli-weight that scales with the linear dimension of the system, $L$. 
An alternative approach is to encode fermions into the emergent fermions of a $\Z_2$ topological order (TO), such as the toric code~\cite{kitaev2003} or Kitaev honeycomb model~\cite{kitaev2006}.
Fermionic hamiltonian simulations with these topological $2d$ F2QEs have been implemented in trapped-ion~\cite{nigmatullin2024experimental} and neutral-atom~\cite{evered2025probing} quatum processors.
In this approach, fermionic bilinears $c_i^\dagger c_j^{\vphantom\dagger}$ are encoded into strings of qubit operators whose length scales with the linear distance between $i$ and $j$, independent of the overall system size.
A toric code on a given lattice naturally encodes one Majorana fermion operator per plaquette of the lattice. Therefore, encoding a lattice model with $N_f$ flavors of complex fermions, requires encoding each site of the fermionic lattice model into an block of $2N_f$ nearby sites of the $\Z_2$ TO.

Early stages of fermionic MERA circuits inherently involve applying unitary operations involving fermionic modes that are close together in the IR, but far apart in the UV~\footnote{Here, we are preparing the state, therefore we view the the RG circuit from IR to UV}.
If one directly encodes the fermions into a $\Z_2$ TO on the UV lattice of qubits, this would result in encoded operators with weight that scales with the system length $O(L)$, resulting in large circuit overhead.
Instead, one can consider encoding the fermionic modes into a $\Z_2$ TO that encodes only the modes that are active at a given RG step, and not the modes that will be injected into by isometries that occur later (deeper in the UV) in the fermionic MERA circuit.
In this way, the length of the gauged fermion strings required to implement an RG step scale only with the block radius, $r$, and not with the system size.
Then, after the unitary disentangling stages are implemented for one RG step, one needs to enlarge the $\Z_2$ TO encoding, by refining the lattice by introducing extra degrees of freedom that can encode the new fermions injected by the isometries of the MERA.
Since the $\Z_2$ TO with lattice spacing $a$ is in the same phase as one with lattice spacing $a/2$, one can refine the lattice of a $\Z_2$ TO to double the degrees of freedom by injecting unentangled qubits, and applying a short-depth quantum circuit to merge them with the $\Z_2$ TO. By having the $\Z_2$ TO expand with the RG, one can ensure that all of the fermionic bilinears involved in the MERA circuit get mapped to qubit strings with weight of order the block radius, $O(r)$, independent of system size. We review the standard Kitaev F2QE encoding in Appendix \ref{F2QE: fermionic_strings}, and provide a detailed proof-of-principle construction for a circuit that implements the  lattice refinement procedure described above in \ref{F2QE: lattice_refinement}.

\subsection{Qubit-efficient implementation with sequential circuits and mid-circuit measurements and reset (MCMR)}
The $2d$ GMERA procedure can be implemented in a qubit-efficient manner by recasting it as a sequential circuit and leveraging mid-circuit measurement and reset (MCMR) techniques \cite{fossfeig2024}, following similar strategies developed for holographic simulation of $1d$ MERA circuit \cite{anad2023holograph}. Rather than implementing the full $2d$ GMERA circuit in parallel, which would require $O(L^2)$ qubits for an $L\times L$ system, we can implement it sequentially by preparing one row of blocks with radius $r$ at a time. This sequential approach dramatically reduces the required number of qubits while preserving the ability to measure all desired observables.

Using MCMR techniques, where physical qubits are measured and reset after each row operation, the total number of required qubits scales as $O(r^2L \log L)$. This includes $O(r^2L)$ physical qubits to encode one row of fermions  with radius $r$ and O(log L) bond qubits to mediate correlations between rows. Details of the implementation of qubit reuse MCMR can be found in Appendix \ref{F2QE: qbit_reuse_MCMR}.

\section{Discussion}
\subsection{Complexity implications}
Our numerical simulations place empirical upper bounds on the quantum resources required to prepare certain types of area-law states. 
Namely, for gapped phases of \emph{correlated} electrons that are adiabatically connected to a non-interacting electron state  the GMERA procedure can be augmented by a constant depth adiabatic evolution~\cite{Hastings_2005} to approximate the interacting ground-state. 
This class of short-range correlated states that can be reliably prepared in this manner include (correlated) trivial and Chern insulators, but not, for example, fractional Chern insulators.
For $2d$ systems with short-range interactions, this adiabatic dressing can be implemented with a fermion-to-qubit encoding with constant overhead, though simulating long-range (e.g. Coulomb) interactions may add additional polynomial (in system size) overhead. 

The results in Fig.~\ref{fig:HaldaneModel} suggest that the error in approximating Gaussian area-law entangled states by $2d$ as a GMERA decays exponentially in the block radius, $r\sim \sqrt{B}$. Equivalently, to achieve a desired target error-per-site, $\epsilon$, one needs to consider $r\gtrsim \log 1/\epsilon$.
The number of two-qubit gates, $G(B)$, required to implement a general Gaussian unitary within a block scales as:
$G_{\rm block}\sim B^2r\sim r^5 \sim \log^5 1/\epsilon$
where the $B^2$ factor comes from the number of fermion rotations to perform the distillation step, and the factor of $r$ comes from F2QE.
Since the GMERA circuit has logarithmic depth, the gate-count for the entire GMERA circuit in system size $L\times L$ is:
$$ G(L,\epsilon) \sim (L/r)^2 \log(L) (B^2r) \approx O((L^2\log L) \log^5(1/\epsilon))$$ 
where the factor of $(L/r)^2$ comes from the number of blocks in each RG step. This scaling applies for the long-range entangled Chern insulator and Dirac semimetal phases.
For a short-range entangled trivial insulator with correlation length $\xi$, one can truncate the MERA RG procedure after $O(\log \xi/B)$ RG steps, removing the $\log L$ factor from the required gate-count.

Finally, we note that with qubit reuse techniques~\cite{fossfeig2024}, the number of qubits, $Q$, required to implement the GMERA circuit can be reduced from $L^2$ (without qubit reuse) to either $O(L\log L \log^2(1/\epsilon))$ (for Chern and DSM) or $O(L\log^2(1/\epsilon))$ for a trivial insulator (truncating GMERA to GPEPS reduce the a $\log L$ factor).

\subsection{Outlook}
We have introduced a $2d$ GMERA procedure to compress a $2d$ Gaussian state, represented by its correlation matrix, into a logarithmic-depth quantum circuit. This approach extends previous work by Fishman and White for $1d$ GMPS and GMERA compression schemes, by introducing a crucial Wannierization procedure to limit the build-up of higher-dimensional correlations that cause a direct extension of $1d$ techniques to fail. The approach can also be considered as a local, circuitization of the recently-introduced continuous time zipper entanglement renormalization (ZER) method~\cite{wong2022zipper}.

By numerically implementing this procedure for a Haldane model on the honeycomb, empirically reveal that the approximation errors in the $2d$ GMERA procedure decrease exponentially in the size of the local blocks used in the procedure.

Compared to other recent approaches like isometric Gaussian fermionic TNS (isoGfTNS)\cite{wu2025alternatinggaussianfermionicisometric}, the $2d$ GMERA offers distinct advantages in circuit depth for large systems. While uni-isoGfTNS requires $O(2L)$ circuit depth and alt-isoGfTNS demands $O(L^2)$ depth, the $2d$ GMERA approach scales as $O((\log L) \log^5(1/\epsilon))$, providing an exponential advantage for large system sizes at fixed target error $\epsilon$. This favorable scaling makes $2d$ GMERA particularly suitable for quantum simulation of large-scale systems.

A number of targets remain for future inquiry. Examples include generalizing the approach to systems without translational invariance (for example to simulate disordered systems or edge states of topological insulators), or to metallic states with Fermi surfaces that violate the area-law entanglement (requiring either block size to scale at least logarithmically in system size, or to consider a more complicated branching MERA type network~\cite{evenbly2014class,haegeman2018rigorous}).

For near-term quantum devices, an important direction is optimizing the GMERA implementation for smaller-scale demonstrations that can be realized with current NISQ hardware. This includes developing architecture-specific compilation strategies - for example, exploiting the all-to-all connectivity of trapped-ion systems and neutral atom arrays to minimize circuit depth and gate count. Other key considerations include exploring error mitigation techniques suited to the GMERA structure, and identifying the minimal system sizes needed to demonstrate key qualitative features of different phases. For example, one could focus on distinguishing between trivial and topological phases using small honeycomb patches that are accessible with current quantum processors, while carefully accounting for hardware connectivity constraints and gate error rates.

Since Gaussian states can be efficiently simulated classically, the main practical utility of an efficient Gaussian fermion state preparation procedure is to serve as an input to other variational or non-variational algorithms for including correlation effects. Highly entangled states in $2d$ or $3d$ is a ripe target for achieving practical quantum advantage.
For gapped systems, correlations can, in-principle be built in by adiabatic dressing as described in the previous section.
For gapless systems, whose gap vanishes as a polynomial in system size, other approaches may be required.
Previous investigations in $1d$ systems indicate that using a Gaussian mean-field/Hartree-Fock as a starting point for variational optimization can dramatically reduce the number of variational parameters required to accurately capture correlated ground-states~\cite{niu2022holographic}. It would be interesting to assess whether the same type of variational improvements can be achieved in $2d$ (a challenging task for classical simulations that may require a quantum processor).

\paragraph*{Acknowledgements -- } We thank Daoheng Niu, Joseph Sullivan conversations, and Yuxuan Zhang for insightful conversations. This
work was supported by the US Department of Energy
DOE DE-SC0022102.
Numerical computations were performed using the Quantum Advanced Research Computing cluster (QuARC) at UBC.

\bibliography{references} 

\appendix{}

\section{Gaussian states and correlation matrices}
\subsection{Gaussian fermion states}
This appendix briefly reviews some notation and formalism for Gaussian fermion states.
Consider a system of fermions with creation operators $\{ \hat{c}^{\dagger}_i \}$,  where the index $i$ can represent the spatial coordinate,  spin, or orbital.  
For simplicity, we consider number-conserving and non-interacting (Gaussian) Hamiltonians of the form: 
\begin{equation}
\hat{H} = \sum_{ij}  \hat{c}^{\dagger}_i h_{ij} \hat{c}_j   \ ,
\end{equation}
where $h_{ij}$ is a hermitian matrix
(the extension to non-number conserving systems is straightforward using a standard Bugoliubov-de-Gennes framework, or working with Majorana representations).

Let $U$ be the unitary matrix that diagonalizes $h$: $(U^\dagger h U)_{ij} = \varepsilon_i \delta_{i,j}$ where $\{\varepsilon_i\}$ are the (single-particle) mode energies. 
Define the eigenmode creation and annihilation operators:
denote 
\begin{equation}
\hat{d}^{\dagger}_{k} = \hat{c}^{\dagger}_{i} U_{ik} \ , 
\hat{d}_{l} = U^{*}_{l j} \hat{c}_{j}.
\end{equation}
In this notation, the ground-state of $\hat{H}$ with chemical potential $\mu$ is:
\begin{equation}
| \Psi \rangle = \prod_{i} (\hat{d}_i^\dagger)^{n_F(\e_i) }| 0 \rangle
\end{equation}
where $|0\>$ is the empty Fock state with no fermions, and $n_F(\e_i) = \theta(\mu-\e_F)$ is the (zero-temperature) Fermi occupation function.

 Wick's theorem implies that all the all correlations can be expressed in terms of combinations of 2-point correlations 
 \begin{align}
C_{ij} = \langle \Psi| \hat{c}^{\dagger}_i \hat{c}_j |\Psi\rangle = \sum_k U_{ik} n_F(\e_k)U^*_{kj} = \iota \iota^\dagger,
\end{align}
where $\iota = Un_F$ is an isometry, and $C$ is a projector onto the occupied orbitals.

The GMPS and GMERA procedures require considering restricted correlations inside a subregion $A$ of the entire system.
The correlation matrix, $C_A$, for subsystem $A$, is simply given by the restriction of the full correlation matrix $C_{i,j}$ to the block with $i,j\in A$.
For a Gaussian state, the reduced density matrix for region $A$ can be expressed as a thermal state of a entanglement (a.k.a. modular) Hamiltonian, $H_A$:
\begin{equation}\label{ch2:rdm}
\hat{\rho}_A = \text{Tr}_{\bar{A}} (|\Psi \rangle \langle \Psi|) = \frac{e^{-\hat{H}_{A}}}{Z}
\end{equation}
where, for a Gaussian state, the entanglement Hamiltonian is also non-interacting: $ \hat{H}_{A} = \sum_{ij \in A}  \hat{c}^{\dagger}_i h^{ij}_{A} \hat{c}_j$, with:
\begin{align}
    C_A &= \frac{1}{\exp(h_A^T)+1}
    \nonumber\\
    h_A &= \log \left(\frac{1-C_A^T}{C_A^T}\right)
\label{eq:haca}
\end{align}
\begin{align}
S_A &= - \text{Tr}( \hat{\rho}_A \log \hat{\rho}_A)
\nonumber\\
&= - \sum_i (\eta_i \log \eta_i +(1-\eta_i)\log(1-\eta_i))
\end{align}
where $0\leq \eta_i \leq 1$ is the eigenvalue of the restricted correlation matrix $C_A$, related to the entanglement energy, $\e^A_i$ ($i^{\rm{th}}$ eigenvalue of $h_A$) by $\eta_i = \frac{1}{e^{\e^A}+1}$. 
Sub-region eigenmodes with $\eta_0\approx 0,1$ ($|\e_{A,i}|\gg 1$) contribute negligibly to $S_A$ and approximately coincide with full,empty eigenmodes of the full correlation matrix that are localized within region $A$.

\section{Expanding Kitaev honeycomb model F2QE}
This Appendix details the expanding Kitaev honeycomb model fermion-to-qubit encoding (F2QE) for 2d fermionic MERA circuits sketched in the main text.
We note that this encoding is not specific to Gaussian states, and may also be used to implement non-Gaussian MERA circuits.

After briefly reviewing the standard method for encoding fermionic bilinears into strings that create emergent fermionic excitations of the Kitaev honeycomb model, we explain how this encoding can be adapted to enable qubit-efficient F2QE. When preparing states using the GMERA circuit from IR to UV, we progressively refine the encoding lattice by halving the lattice spacing at each RG step to accommodate new fermion modes as they are injected by the isometries. Additionally, we show how recasting the GMERA as a sequential circuit using mid-circuit measurement and reset techniques reduce the required number of qubits.

\subsection{Fermionic strings in Kitaev honeycomb model} \label{F2QE: fermionic_strings}
\begin{figure*}[t]
    \begin{center}
\includegraphics[width=\textwidth]{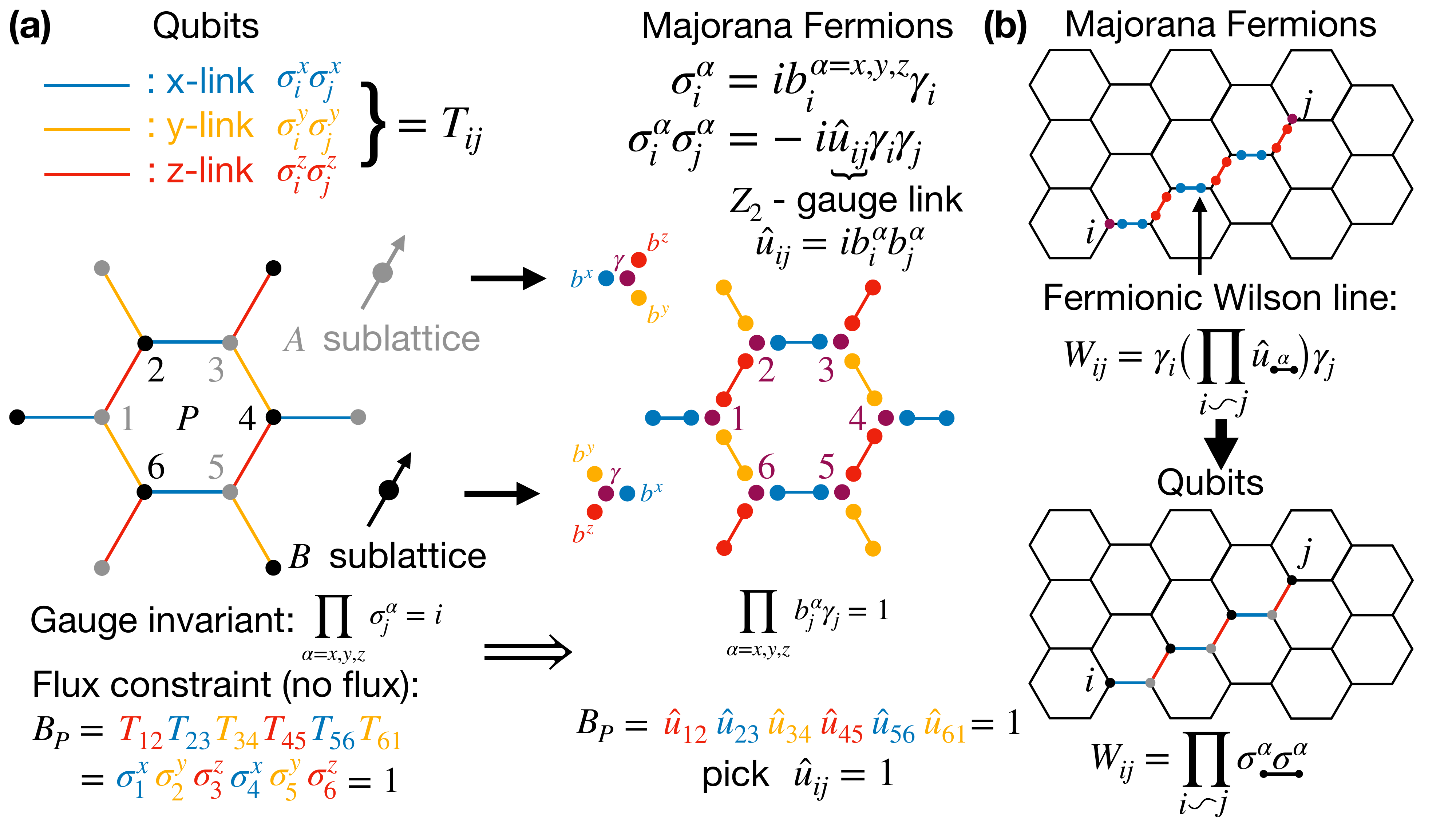} 
    \end{center}
    \caption{\textbf{Fermion-to-qubit encoding using the Kitaev honeycomb model.} (a) Mapping between qubit operators and Majorana fermions in the Kitaev honeycomb model. Left: The honeycomb lattice with x-links (blue), y-links (yellow), and z-links (red) connecting sites on the A (gray) and B (black) sublattices. Each site hosts a qubit, and the corresponding Pauli operators form link operators $T_{ij}$. Right: In the Majorana representation, each site hosts four Majorana fermions (one $c$ and three $b^{\alpha}$), with gauge constraints relating them. The plaquette operator $B_P$ enforces the zero-flux constraint essential for the topological encoding. (b) Construction of fermionic Wilson lines between distant sites. A string of link operators connecting sites $i$ and $j$ encodes a fermionic bilinear operator that properly accounts for fermionic statistics. This encoding enables implementing fermionic operations with strings whose length scales with the geometric distance between sites rather than with the system size, which would be required for a Jordan-Wigner encoding.}
    \label{fig:kitaev-F2Q}
\end{figure*}

The Kitaev honeycomb model~\cite{kitaev2006} provides an efficient fermion-to-qubit encoding (F2QE) for implementing fermionic operations in 2D systems (see \cite{evered2025probing} for a recent experimental implementation of this encoding).
Unlike the Jordan-Wigner transformation which produces strings that scale with system size, this encoding maps fermionic operators to strings of Pauli operators with length scaling only with the geometric distance between sites.

The Kitaev honeycomb model is defined on a honeycomb lattice with qubits at each vertex, divided into two sublattices, A and B. The lattice features three types of links $x$-links (blue), $y$-links (yellow), and $z$-links (red) which interactions occur. The Hamiltonian is given by:
\begin{equation}
    H = -J_x\sum_{\text{x-links}}\sigma^x_i\sigma^x_j-J_y\sum_{\text{y-links}}\sigma^y_i\sigma^y_j-J_z\sum_{\text{z-links}}\sigma^z_i\sigma^z_j
\end{equation}
where $\sigma_i^{\alpha}$ $(\alpha=x,y,z)$ are Pauli operators acting on qubit $i$, and $J_x,J_y,J_z$ are coupling constants. 

In the Majorana representation, each qubit at site $i$ is expressed using four Majorana ``partons": $b_{i}^x,b_{i}^y,b_{i}^z,\gamma_i$. These operators are self-adjoint $((b^{\alpha}_i)^{\dagger} = b^{\alpha}_i, \gamma_i^{\dagger} = \gamma_i)$, square to the identity $(b^{\alpha}_i)^2 = \gamma_i=1$, and obey anticommutation relations: $\{b_i^{\alpha},b_j^{\beta} \} = 2\delta_{ij}\delta_{\alpha\beta},\{\gamma_i\gamma_j \}=2\delta_{i,j} = \{b_i^{\alpha},\gamma_i \}=0$.
The Pauli operators are constructed as:
\begin{equation}
    \sigma^{\alpha}_i = ib^{\alpha}_i\gamma_i, \ (\alpha=x,y,z)
\end{equation}
To restrict the 4-dimensional Majorana Hilbert space to the physical spin-1/2 one, a local $\Z_2$ gauge constraintL
\begin{equation}
    D_i = b_i^xb_i^yb_i^z\gamma_i = 1.
\end{equation}
is enforced at each site, $i$.
The degrees of freedom 
This representation reveals emergent fermionic excitations, which we use to define complex fermions.

On an $\alpha$-type link, $ij$, the $b^\alpha$ fermions can be organized into $\Z_2$ gauge connections:
\begin{align}
    \hat{u}_{ij} = ib_i^\alpha b^\alpha_j.
\end{align}
Products of $\hat{u}_{ij}$ around each hexagonal plaquette, $B_P = \prod_{ij \in P} \hat{u}_{ij}$ (or around any closed loop of links) are gauge invariant (commute with the $D_i$'s) and can be interpreted as measuring the $\Z_2$ gauge flux through the plaquette $P$.
The remaining fermions $\gamma_i$ are viewed as emergent fermionic matter that are minimally coupled to the gauge field $\hat{u}_{ij}$.

The gauge-invariant Wilson line operators: 
\begin{align}
W_{i,j}=
i\gamma_i \prod_{kl \in \Gamma_{i,j}} \hat{u}_{kl} \gamma_j,
\end{align} 
can be written as Pauli strings of the physical spin operators along the (oriented) path $\Gamma$ connecting site $i$ to $j$. 

In the subspace with vanishing  gauge flux (``flat" gauge connection), stabilized by $B_P = +1 \forall P$, we are free to choose a gauge where $\hat{u}_{ij}=+1 \forall ij$, and the Wilson line operators manifestly have the same algebraic structure as that of ordinary (un-gauged) fermion bilinears, $\{\gamma_i\gamma_j\}$, and hence represent an F2QE encoding. Since, for a flat $\Z_2$ gauge field, the Wilson strings are independent of path, the most efficient such F2QE encoding corresponds to simply choosing $W_{i,j}$ to be implemented by the shortest path from $i$ to $j$.

These Majorana operators $\gamma_i$ can be paired to represent complex fermions. For example, pairing the neighboring $\gamma$ modes along $z$-links connecting sites $i$ (on sublattice A) and $j$ (on sublattice B) defines the complex fermion operators:
\begin{equation}
    c_{ij}=\frac{\gamma_i+i\hat{u}_{ij}\gamma_j}{2}\ \ \  \text{ and }\ \ \ c^{\dagger}_{ij}=\frac{\gamma_i-i\hat{u}_{ij}\gamma_j}{2}
\end{equation}
with corresponding number operator:
\begin{equation}
    n_{ij} = c_{ij}^{\dagger}c_{ij} = \frac{1}{4}(\gamma_i-i\gamma_j)(\gamma_i+i\gamma_j)=\frac{1}{2}(1+i\gamma_i\hat{u}_{ij}\gamma_j)
\end{equation}

For instance, encoding a fermionic rotation $e^{i\theta c^{\dagger}_{ij}c_{kl}}$ involving modes $c{\dagger}_{ij}, c_{kl}$ where $z$-bonds $(i,j)$ and $(k,l)$ are separated by distance $r$, within radius $r$ is implemented by performing a rotation $e^{i \theta P}$ by the encoding Pauli string of spin operators with Pauli-weight $O(r)$.

\subsection{Kitaev honeycomb model: Lattice Refinement Circuit} \label{F2QE: lattice_refinement}
\begin{figure*}[t]
    \begin{center}
\includegraphics[width=\textwidth]{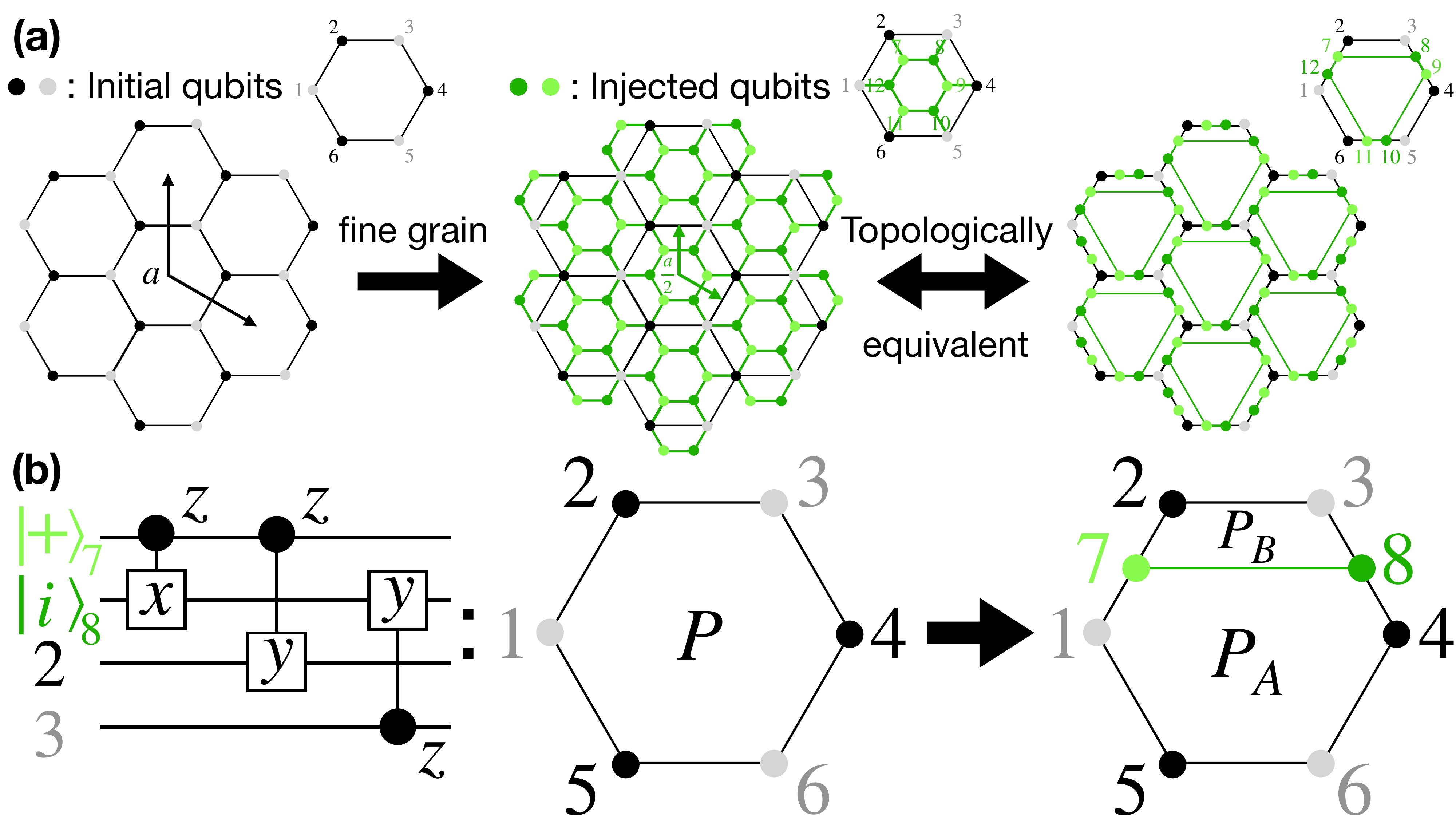} 
    \end{center}
    \caption{{\bf Lattice refinement process for Kitaev honeycomb encoding.} (a) Starting from an initial honeycomb lattice with lattice spacing $a$ (black and gray vertices), we inject additional qubits (green vertices) to obtain a refined lattice with spacing $a/2$. This refined lattice can be rearranged into a topologically equivalent structure as shown on the right. (b) Example of injecting a pair of qubits (7,8). The original plaquette operator $B_{P}$ splits into $B_{P_A}$ and $B_{P_B}$ in the refined lattice as shown on the right. The Clifford circuit on the left would first initialize the qubits to satisfy the stabilizer conditions $\sigma^x_7\sigma^x_8 = +1$ (fixing charge sector) and $\sigma^z_7\sigma^y_8 = +1$ (fixing flux sector), then map the stabilizer $\sigma^z_7\sigma^y_8$ to $B_{P_B}=\sigma_7^z\sigma_2^y\sigma^z_3\sigma_8^y$ while preseving the original plaquette stabilizer $B_P = \sigma_1^x\sigma_2^y\sigma^3_z\sigma^4_x\sigma^5_y\sigma^6_z$.
    Therefore, if $B_{P} = 1$ initially, both $B_{P_A}$ and $B_{P_B}$ equal 1 after the circuit. Similarly, when injecting qubits (9,10), they would be initialized to satisfy $\sigma^z_9\sigma^z_{10} = 1$ and $\sigma^y_9\sigma^z_{10} = 1$, and for qubits (11,12), the initialization would be $\sigma^y_{11}\sigma^y_{12} = 1$ and $\sigma^z_{11}\sigma^x_{12} = 1$, following the bond orientations.}
    \label{fig:lattice-refinement}
\end{figure*}
Attempting to directly implement a fermionic MERA circuit using a $2d$ F2QE encoding would result in large circuit overheads since early (IR) steps of the MERA correspond to entangling distant (with the UV metric) fermionic modes, which needs to be implemented via high-weight Pauli rotation operations.
To evade this, we propose to have the honeycomb lattice defining the $2d$ Kitaev F2QE scale along with the MERA circuit, by injecting additional un-entangled fermionic modes into the lattice with each MERA step (with circuit time proceeding from IR to UV).
In this Appendix, we describe a a concrete realization of a short-depth quantum circuit that maps a state of a 2d Kitaev honeycomb model with lattice spacing $a$ to one with lattice spacing $a/2$, in a manner that preserves the gauge electric charge and flux configurations at scale $a$, which we refer to as a lattice-refinement of the original state of Kitaev honeycomb model. 
This is accomplished by injecting additional qubits encoding additional (unentangled) fermionic modes as depicted in Fig.~\ref{fig:lattice-refinement}a.
We design an injection circuit to both 1) maintain the state of the initial fermionic degrees of freedom, and 2) preserve the flat gauge field condition $B_P=+1$ as the lattice is refined.

The lattice refinement in Fig.~\ref{fig:lattice-refinement}a, can be decomposed into a few layers of more elementary plaquette refinement operations, such as the one shown in Fig.~\ref{fig:lattice-refinement}b.
For concreteness, we focus on the specific example of injecting a single fermionic mode, represented by two injected qubits $7,8$ connected along a new $x$-type bond, into a plaquette with initial qubits labeled $1,2,\dots 6$.
After injection the plaquette, $P$, is split into two plaquettes, $P_A$ and $P_B$. 
The Clifford circuit shown in Fig.~\ref{fig:lattice-refinement}b, would first initialize the injected qubits in simultaneous $+1$ eigenstate of the operators: $\sigma^x_7\sigma^x_8 = i\gamma_7u_{7,8}\gamma_8$, and $\sigma^z_7\sigma^y_8$, then entangles the new fermion mode on sites $7,8$ into the initial Honeycomb encoding, mapping the initial stabilizer $\sigma^z_7\sigma^8_y=+1$ into one of the new plaquette stabilizers: $B_{P_B} = \sigma_7^z\sigma_2^y\sigma^z_3\sigma_8^y=+1$, while preserving the original plaquette stabilizer $B_P = \sigma_1^x\sigma_2^y\sigma^3_z\sigma^4_x\sigma^5_y\sigma^6_z=+1$.

The injection procedure for other orientations of bonds proceeds similarly but the Pauli basis cyclically permuted according to the appropriate bond orientations in the Honeycomb. For example, referring to the site numbering indicated in Fig.~\ref{fig:lattice-refinement}a, injected qubits $9,10$ are injected by a similar circuit as $7,8$ but with $x\rightarrow z\rightarrow y \rightarrow x$, and qubits $11,12$ are injected by cycling $x\rightarrow z \rightarrow y \rightarrow x$.

The entire refinement shown in Fig.~\ref{fig:lattice-refinement}a can be implemented in 3 circuit layers consisting of injecting bonds like (7,8) on all plaquettes followed by injecting (9,10) and then (11,12) type bonds.
This sequence results in a constant depth circuit that implements the desired lattice refinement.

\subsection{Qubit-efficient implementation with sequential circuits and MCMR}\label{F2QE: qbit_reuse_MCMR}
\begin{figure*}[t]
    \begin{center}
\includegraphics[width=\textwidth]{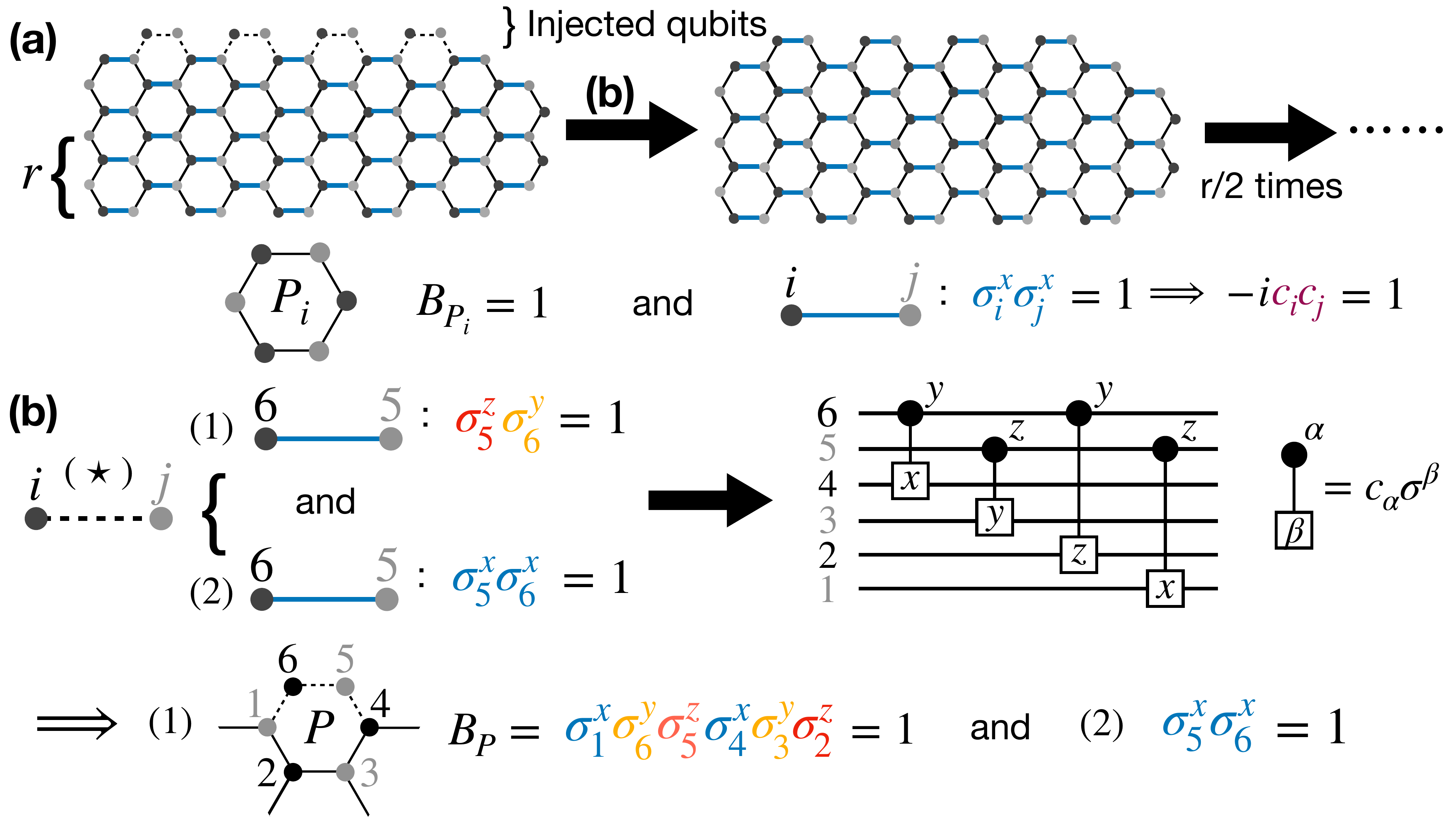} 
    \end{center}
    \caption{\textbf{Sequential preparation of fermions in the GMERA process with expanding encoding.} 
    (a) The honeycomb lattice showing both original qubits (black/gray) and injected qubits that form the expanding encoding. Parameter $r$ indicates the radius of local regions used in the initial GMERA step, determining the scope when identifying locally entangled modes. Each plaquette $P_i$ requires the flux constraint $B_{P_i} = 1$ (original qubits), while x-links between sites $i$ and $j$ establish the relation $\sigma_i^x\sigma_j^x = 1$, corresponding to the magnetic pairings $-ic_ic_j = 1$. To expand the system for sequential preparation, we must repeat the process indicated in (b) $r/2$ times, where $r/2$ represents the distance between shifted regions in each disentangling step. (b) The sequential preparation process where initially disentangled qubits (connected by dashed lines) are initialized with two key stabilizer conditions: (1) $\sigma^z_5\sigma^y_6 = 1$ and (2) $\sigma^x_5\sigma^x_6 = 1$. When transformed through the stabilizer circuit shown in the middle, the first condition $\sigma^z_5\sigma^y_6 = 1$ directly transforms into the flux constraint $B_P = \sigma^x_1\sigma^y_6\sigma^z_5\sigma^x_4\sigma^y_3\sigma^z_2 = 1$, while the second condition $\sigma^x_5\sigma^x_6 = 1$ commutes with the circuit and remains unchanged. This initialization strategy ensures the preservation of both the flux and charge constraints, maintaining the $Z_2$ topological order throughout the expanding encoding process.}
    \label{fig:sequential_prep}
\end{figure*}

If our primary goal is to measure physical observables rather than preparing the entire ground state of a 2D system, we can significantly reduce qubit resources by implementing the 2D GMERA procedure as a sequential circuit with mid-circuit measurement and reset (MCMR).

In the standard parallel implementation, the full GMERA circuit would require $O(L^2)$ qubits for an $L \times L$ system, making it impractical for large system sizes on near-term quantum hardware. Instead, we can implement the GMERA circuit sequentially by preparing one row of blocks with radius $r$ at a time. After applying the operations for a given row and measuring the necessary observables, we can reset and reuse those physical qubits for the next row, while keeping only the bond qubits that mediate correlations between different rows.

This sequential approach dramatically reduces the required qubit count while preserving our ability to measure all desired observables. The total number of qubits needed scales as:
\begin{equation}
O(r^2L \log L) = O(\log^2(1/\epsilon)L \log L)
\end{equation}

This includes $O(r^2L)$ physical qubits to encode one row of sites with block radius $r$ and $O(\log L)$ bond qubits to mediate correlations across different RG levels. The $\log L$ factor arises from the hierarchical structure of the GMERA, where each RG step involves iterating over the system at increasingly larger length scales (with the lattice refinement circuit introduce in \ref{F2QE: lattice_refinement}).

The sequential implementation with MCMR works as follows:
\begin{enumerate}
\item Initialize the bond qubits in a state representing the configuration at the IR (infrared) level
\item Prepare the first row of blocks according to the GMERA procedure
\item Measure any desired observables on this row
\item Reset the physical qubits while preserving the bond qubits
\item Prepare the second row using the bond qubits to mediate correlations with the first row
\item Continue this process for all rows in the system
\end{enumerate}

Additionally, when implementing the expanding Kitaev honeycomb model encoding, we need to ensure that the flux constraints are properly maintained during the sequential preparation.

For the injected qubits $(5,6)$ shown in Fig. \ref{fig:sequential_prep}, we impose two initial constraints:
\begin{enumerate}
\item $\sigma^z_5\sigma^y_6 = 1$ (flux constraint)
\item $\sigma^x_5\sigma^x_6 = 1$ (charge conservation)
\end{enumerate}

When the  $\sigma^x_5\sigma^x_6 = 1$ constraint passes through the stabilizer circuit, it transforms into:
\begin{equation}
\sigma^x_1\sigma^y_6\sigma^z_5\sigma^x_4\sigma^y_3\sigma^z_2 = 1
\end{equation}

This ensures that the flux constraint $B_P = 1$ is preserved, maintaining the topological properties of the $Z_2$ encoding as we expand the lattice to accommodate new fermion modes at each RG step.

The sequential circuit preparation and expanding Kitaev encoding can be combined to sequentially prepare a 2d GMERA following a straightforward generalization of the $1d$ sequential MERA construction explained in~\cite{aanand2022a}.


\end{document}